%% file: figures/main.tex
\documentclass[aps,prl,twocolumn,superscriptaddress,showkeys]{revtex4-1}
\usepackage{graphicx}
\usepackage{amsmath}
\usepackage{amssymb}
\usepackage[colorlinks=true,citecolor=blue,urlcolor=blue,linkcolor=blue]{hyperref}
\usepackage{bm}
\usepackage{physics}
\usepackage{xcolor}
\usepackage{placeins}

\usepackage{tikz}
\usepackage{pgfplots}
\usetikzlibrary{positioning,shadows,arrows,calc,shapes,decorations.pathreplacing,matrix,shapes.geometric,patterns,decorations.markings,math}

\tikzstyle{startstop} = [rectangle, rounded corners, minimum width=3cm, minimum height=1cm,text centered, draw=black, fill=red!30]
\tikzstyle{io} = [trapezium, trapezium left angle=70, trapezium right angle=110, minimum width=3cm, minimum height=1cm, text centered, draw=black, fill=blue!30]
\tikzstyle{process} = [rectangle, minimum width=3cm, minimum height=1cm, text centered, draw=black, fill=orange!30]
\tikzstyle{decision} = [diamond, minimum width=3cm, minimum height=1cm, text centered, draw=black, fill=green!30]
\tikzstyle{arrow} = [thick,->,>=stealth]
\tikzstyle{matrix} = [rectangle, minimum width=2.5cm, minimum height=1cm, text centered, text width=2.5cm, draw=black, fill=yellow!30]

\definecolor{soothingblue}{RGB}{204,229,255} 
\definecolor{soothingred}{rgb}{1, 0.6, 0.6} 
\definecolor{soothinggreen}{RGB}{180,255,180} 
\definecolor{darkerred}{rgb}{0.6,0,0.15} 
\definecolor{darkerblue}{rgb}{0.2,0.2,0.7} 
\definecolor{darkergreen}{rgb}{0.2,0.6,0.15} 
\usepackage{standalone}

\usepackage{diagbox}
\usepackage{makecell}
\usepackage{soul}
\usepackage{multirow}
\usepackage{colortbl}
\makeatletter
\newcommand{\setCustomStrut}{%
  \setbox\@arstrutbox=\hbox{\vrule height 3.3ex depth 1.7ex width 0pt}%
}
\makeatother

\begin{document}

\title{Surface-Code Thresholds and Qubit Footprints in Shuttling-Based Spin-Qubit Railways}

\author{Arun John {Moncy}}
\affiliation{Donostia International Physics Center, 20018 San Sebastian, Spain.}
\affiliation{Department of Basic Sciences, Tecnun - University of Navarra, 20018 San Sebastian, Spain.}
\author{Reza Dastbasteh}
\affiliation{Department of Basic Sciences, Tecnun - University of Navarra, 20018 San Sebastian, Spain.}
\author{Josu {Etxezarreta Martinez}}
\affiliation{Department of Basic Sciences, Tecnun - University of Navarra, 20018 San Sebastian, Spain.}
\author{Ryo Nagai}
\affiliation{Research and Development Group, Hitachi, Ltd., Kokubunji, Tokyo 185-8601, Japan}
\author{Pedro M. Crespo}
\affiliation{Department of Basic Sciences, Tecnun - University of Navarra, 20018 San Sebastian, Spain.}
\author{Normann Mertig}
\affiliation{Hitachi Cambridge Laboratory, J. J. Thomson Avenue, Cambridge, CB3 0HE, United Kingdom.}
\author{Charles Smith}
\affiliation{Hitachi Cambridge Laboratory, J. J. Thomson Avenue, Cambridge, CB3 0HE, United Kingdom.}
\author{Ruben M. Otxoa}
\affiliation{Hitachi Cambridge Laboratory, J. J. Thomson Avenue, Cambridge, CB3 0HE, United Kingdom.}
\date{\today}

\begin{abstract}
We present a fault-tolerant mapping of rotated surface codes onto a $2\times N$ silicon spin-qubit railway architecture, utilizing electron shuttling to resolve the wiring fan-out bottleneck. Employing circuit-level noise modeling, we evaluate threshold performances across various noise biases. We demonstrate that shuttling check qubits instead of data qubits fundamentally improves system thresholds. Crucially, under a noise model biased towards dephasing for spin-qubit shuttling, the non-CSS XZZX surface code outperforms standard CSS variants. By tailoring the topological code to this specific inherent  bias, we show that the Megaquop footprint is achievable with a distance 7 code requiring a $p=10^{-3}$ physical error rate, highlighting a pathway for substantial hardware reductions in early fault-tolerant quantum processors.
\end{abstract}

% PACS is formally deprecated; APS now uses PhySH, but keywords remain useful.
\keywords{Quantum error correction, Surface codes, Silicon spin qubits, Coherent electron shuttling, Biased noise, Fast syndrome extraction}

\maketitle

\section{I. Introduction}
Realizing the full potential of quantum computing requires the implementation of fault-tolerant protocols to suppress physical error rates to arbitrarily low levels. Among the various quantum error correction (QEC) schemes, the surface code is widely considered the leading candidate for near-term implementation due to its high error threshold and its requirement for only nearest-neighbor interactions on a two-dimensional (2D) lattice. However, while 2D qubit layouts are the natural native geometry for the surface code, fabricating dense, large-scale 2D arrays presents significant engineering problems. In platform candidates such as semiconductor quantum dots, creating the necessary wiring fan-out to control interior qubits without sacrificing coherence or density remains a formidable system-level challenge \cite{li2018crossbar, franke2019rent}.

Silicon spin qubits have emerged as a compelling platform for scalable quantum computing, offering long coherence times and compatibility with industrial CMOS manufacturing processes \cite{gonzalez2021scaling, imec2024industrial, koch2025industrial}. The leveraging of advanced semiconductor fabrication techniques suggests a pathway to millions of qubits, analogous to classical processors. However the wiring bottleneck is particularly problematic in this platform, where gate pitches are on the order of 50-100 nm. Standard 2D topologies and dense crossbar networks \cite{Veldhorst2017_CMOSarchitecture, Li2018_crossbar, Brousse2022_dense2D} effectively crowd out the necessary control lines, forcing a trade-off between qubit density and control fidelity. To circumvent these interconnect bottlenecks, architectures with lower dimensionality have been proposed as a necessary evolution toward fault tolerance.

A promising solution is the use of coherent electron shuttling to generate effective connectivity. By physically transporting electron's spin between interaction zones, the requirement for a static nearest-neighbor grid is relaxed. Recent experimental milestones have demonstrated the viability of this approach, with coherent spin transport fidelities exceeding 99.9\% in silicon devices \cite{yoneda2021coherent, mills2019shuttling, zwerver2023shuttling, desmet2024high, volmer2024mapping, struck2024spin}. This capability has promoted the proposal of new architectural paradigms, such as the ``SpinBus" \cite{kunne2023spinbus, ginzel2024scalable} and ``Pipeline'' processors \cite{patomaki2023pipeline}, which rely on shuttling to interconnect sparse qubits. Similarly, the SpinHex architecture mitigates crosstalk and enhances connectivity through the use of multi-electron couplers \cite{otxoa2025spinhex}. In these frameworks, the 2D surface code is not mapped onto a static 2D lattice, but rather onto a dynamic graph where physical qubits are brought together temporally to perform entangling gate logical operations. Consequently, to fully exploit these dynamical architectures, the development of high-fidelity shuttling protocols is paramount to minimize decoherence and operational errors during transport \cite{ginzel2020spin, kanaar2022nonadiabatic}.

In this work, we investigate the mapping of the surface code onto a $2 \times N$ ``railway" architecture using silicon spin qubits. This bilinear geometry represents a practical compromise: it significantly alleviates the wiring fan-out problem by allowing control lines to access qubits from the edges of the array \cite{li2025trilinear, Escofet2025Quantum, Chadwick2025_SNAQ}, yet it retains sufficient connectivity to embed error correcting codes when augmented with spin-qubit transport. Unlike static 2D layouts, the $2 \times N$ railway architecture allows for reconfigurable connectivity graphs, limited only by the latency and fidelity of the shuttling operations. We develop a protocol for syndrome extraction that enforces commutation constraints and actively prevents hook errors that degrades the performance of the code.
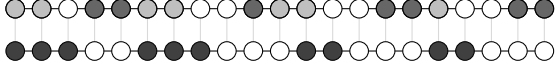
\begin{figure}%[t]
    \centering
    \input{figures/2N}
    \caption{Schematic representation mapping the 2D rotated surface code onto the $2 \times N$ railway architecture.
    Data (black) and check (light and dark gray) qubits are designated strictly to separate, parallel linear `rails'. 
    The necessary connectivity for syndrome extraction and stabilizer measurement is achieved dynamically via coherent electron shuttling operations along these rails. 
    The different circles represent different quantum dot components: white circles are empty quantum dots, black circles represent data qubits, and blue and red circles are check qubits for detecting Z or X error.}
    \label{fig:2N}
\end{figure}
Additionally, we develop a toy noise model that captures the essence of electron shuttling in the presence of micromagnets and spin-orbit coupling effects. While micromagnets are essential for enabling high-fidelity electric-dipole spin resonance (EDSR) gates and addressability in spin qubit devices \cite{tokura2006coherent, pioro2008electrically, yoneda2015robust, klemt2023electrical, gresch2021low}, they introduce significant magnetic field gradients along the transport lines. Our model specifically accounts for the phase accumulation and potential decoherence arising from traversing these gradients, as well as the impact of path-dependent noise. Finally, we examine the impact of spin-orbit coupling during shuttling, which due to relativistic effects subjects the electron to an effective magnetic field that compromises the preservation of coherent states \cite{Bosco2023_shuttlingSOC, Ademi2025_shuttling}. By incorporating these physical constraints into our error model, we provide a realistic assessment of the logical error rates achievable in a $2 \times N$ architecture.

Using circuit-level noise modeling tailored to shuttling-based syndrome extraction \cite{Siegel2024Towards}, we quantify how code choice and qubit-motion strategy reshape the logical threshold when the physical noise is asymmetric across Pauli channels. In particular, we benchmark a standard CSS surface-code layout \cite{fowler2012surface, litinski2019game} against a non-CSS alternative (XZZX) \cite{bonilla2021xzzx,Etxezarreta2026leveraging} under the same gate set, idling, and shuttling conditions \cite{tomita2014low}. Furthermore, we explicitly distinguish two operational modes: moving data qubits \cite{patomaki2023pipeline} versus moving check qubits around stationary data \cite{kunne2023spinbus}, demonstrating that moving the check qubits systematically yields to high fault-tolerant thresholds. We continue our study by numerically exploring the mega- and gigaquop footprints for two experimentally relevant physical error rates, p = 0.003 and p = 0.001. We observe that as a result of the bias tailoring, footprint reductions ranging from 33\% up to 75\% are obtained for all relevant quantum operation regimes.

\section{II. Model}

In recent system-level analyses of spin-qubit architectures, shuttling noise is frequently modeled as an additive, pure dephasing error channel per transport step (e.g., $p_X = p_Y =0; p_Z = p_{\text{shuttle}}$) \cite{Siegel2024Towards}. While these assumptions provide useful insights for a baseline error threshold, they disregard underlying geometric and physical constraints of the spin-qubit architecture. More specifically, it is expected that for realistic trajectory fluctuations through micromagnet gradients and anisotropic Spin Orbit Coupling (SOC) (e.g., interplay between Rashba and Dresselhaus terms) inevitably break this ideal noise channel, resulting in finite biased noise. To accurately capture this bias and evaluate its impact on the code performance, we derive a comprehensive, asymmetric Pauli channel from the underlying physical Hamiltonian, as we detail below. 
\subsection{Composite incoherent channel for a shuttling segment}
\label{sec:coherent_shuttling}
We consider a single-electron spin qubit transported between gate-defined sites while subject to a static applied magnetic field and a spatially inhomogeneous micromagnet (MM) field. In addition, transport may induce effective fields mediated by spin-orbit coupling (SOC). Over a shuttling segment of duration $\tau$, the spin dynamics are described by the time-dependent Hamiltonian
\begin{equation}
H(t)=\frac{g\mu_B}{2}\Big[\mathbf{B}_{\rm ext}+\mathbf{B}_{\rm MM}(\mathbf{r}(t))+\mathbf{B}_{\delta\hat{v}}(t)\Big]\cdot\boldsymbol{\sigma},
\label{eq:H_total}
\end{equation}
where $g$ is the Land\'{e} g-factor, $\mu_B$ is the Bohr magneton, $\boldsymbol{\sigma}$ is the vector of Pauli matrices, $\mathbf{B}_{\rm ext}$ is the externally applied field, $\mathbf{B}_{\rm MM}$ is the micromagnet field evaluated along the trajectory $\mathbf{r}(t)$, and $\mathbf{B}_{\delta\hat{v}}(t)$ is an effective SOC field whose magnitude and direction depend on the instantaneous momentum and local electrostatic environment. The evolution of the spin state under these time-varying fields is described by the unitary propagator, which generates the coherent mapping of the wavefunction from $t=0$ to $t=\tau$ via the time-ordered exponential. To formalize the effect of transport on the qubit, we describe the spin evolution during a single shuttling segment as a driven, time-dependent unitary process. The time dependence enters through the explicit dependence of the local magnetic environment on the instantaneous electron position $\mathbf{r}(t)$, as well as through any velocity- or momentum-dependent contributions such as SOC. Consequently, even for an otherwise idle qubit, shuttling implements a deterministic evolution that must be accounted for at the circuit level. We therefore characterize the action of one shuttling event by the propagator generated by the total instantaneous Hamiltonian in Eq.~(1). Hence the corresponding time-ordered unitary evolution operator over a time duration $\tau$ is given by:
\begin{equation}
U_{\rm shuttle}=\mathcal{T}\exp\!\left[-\frac{i}{\hbar}\int_0^\tau H(t)\,dt\right],
\label{eq:U_shuttle}
\end{equation}
with $\mathcal{T}$ denoting time ordering and $\hbar$ the reduced Planck constant. To make explicit the dependence of the spin evolution on transport, we expand the MM field about a path $\mathbf{r}_0(t)$ and small trajectory deviations $\delta\mathbf{r}(t)=\mathbf{r}(t)-\mathbf{r}_0(t)$, while also incorporating intrinsic magnetic field fluctuations $\delta\mathbf{B}(t)$. To first order,
\begin{equation}
\begin{split}
B_{{\rm MM},i}(\mathbf{r}(t)) \simeq B_{{\rm MM},i}(\mathbf{r}_0(t))&+\\ \sum_{j\in\{x,y,z\}} b_{ij}(\mathbf{r}_0(t))\,\delta r_j(t) + \delta B_i(t), \\
\end{split}
\label{eq:MM_linear}
\end{equation}
where $b_{ij}(\mathbf{r}) \equiv \partial B_{{\rm MM},i}/\partial r_j$ is the magnetic field gradient tensor, and $\delta B_i(t)$ captures time-dependent stochastic magnetic noise. Even when $\delta\mathbf{r}(t)=0$ and $\delta\mathbf{B}(t)=0$, the deterministic term $\mathbf{B}_{\rm MM}(\mathbf{r}_0(t))$ generally differs from the static calibration point and produces an additional unitary rotation during transport. This effect is controlled by the known MM field profile and by the timing of the shuttling waveform, and therefore constitutes a coherent error mechanism.
Concurrent with these magnetic gradients, the spin state is also subject to SOC during transport. In the standard Rashba--Dresselhaus description of a 2D electron system, the SOC Hamiltonian is linear in momentum and can be explicitly written as
\begin{equation}
\begin{split}
H_{\delta\hat{v}}(t) &= \alpha_R (\sigma_x p_y - \sigma_y p_x) + \beta_D (\sigma_x p_x - \sigma_y p_y) \\
&\equiv \frac{g\mu_B}{2}\,\mathbf{B}_{\delta\hat{v}}(t)\cdot\boldsymbol{\sigma},
\end{split}
\label{eq:H_SO}
\end{equation}
where $\alpha_R$ and $\beta_D$ are the Rashba and Dresselhaus SOC strengths, respectively. To clearly identify the coherent and incoherent contributions from transport, we express the components of the instantaneous linear momentum as $p_i(t)=m^*(v_i(t) + \delta v_i(t))$, where $m^*$ is the effective mass, $v_i(t)$ is the deterministic shuttling velocity along spatial axis $i$, and $\delta v_i(t)$ represents stochastic velocity fluctuations. Accordingly, the effective SOC field decomposes into a coherent part $\mathbf{B}_{\delta\hat{v}}^{(0)}(t)$ proportional to the deterministic velocity and a stochastic part $\delta\mathbf{B}_{\delta\hat{v}}(t)$ driven by $\delta v_i(t)$. For reproducible shuttling event, the coherent component generates a repeatable spin rotation, which naturally combines with the deterministic MM contribution to form the total coherent error.
To formalize this, we decompose the total Hamiltonian into:
\begin{equation}
H(t)=H_0(t)+\delta H_{\rm coh}(t)+\delta H_{\rm stoch}(t),
\label{eq:H_decomp}
\end{equation}
where $H_0(t)$ defines the target evolution in the chosen computational frame (typically set by $\mathbf{B}_{\rm ext}$ and the static calibration at each site), $\delta H_{\rm coh}(t)$ collects reproducible contributions arising from the deterministic sampling of $\mathbf{B}_{\rm MM}(\mathbf{r}_0(t))$ and the coherent SOC field $\mathbf{B}_{\delta\hat{v}}^{(0)}(t)$, and $\delta H_{\rm stoch}(t)$ accounts for shot-to-shot variations (e.g., residual trajectory fluctuations, field drift, and velocity fluctuations $\delta v_i(t)$). The coherent shuttling unitary is then
\begin{equation}
U_{\rm coh}=\mathcal{T}\exp\!\left[-\frac{i}{\hbar}\int_0^\tau \delta H_{\rm coh}(t)\,dt\right],
\label{eq:U_coh}
\end{equation}
and the full evolution may be written as $U_{\rm shuttle}=U_0\,U_{\rm coh}\,U_{\rm stoch}$, with $U_0=\mathcal{T}\exp[-\tfrac{i}{\hbar}\int_0^\tau H_0(t)\,dt]$.
In many operating regimes relevant to silicon spin qubits, $\mathbf{B}_{\rm ext}$ defines a dominant quantization axis and the coherent shuttling effect can be well-approximated as an additional rotation by a fixed angle around a given axis. When the dominant systematic contribution is longitudinal, it can be compensated by a virtual-frame update without any additional physical operation. More generally, if $U_{\rm coh}$ corresponds to a reproducible rotation about a tilted axis, its effect may be removed by applying a calibrated inverse operation. The remaining contribution is associated with $\delta H_{\rm stoch}(t)$, which leads to incoherent errors and is treated separately in the stochastic noise model presented next.

\vspace{-1.5em}
\section{III. Incoherent shuttling}
\label{sec:incoherent_shuttling}

Before detailing the physical mechanisms of noise, we emphasize that the analytical framework developed in this section is intentionally architecture-agnostic. As will be shown, the resulting error map is sufficiently general to accommodate strong symmetry-breaking scenarios such as extreme phase-error bias ($Z$-biased noise) or strong transversality ($X$-biased noise) as well as highly symmetric depolarizing noise, depending solely on the specific geometrical constraints and operating parameters of the underlying shuttling path. 

It is expected that shuttling is accompanied by shot-to-shot variations in the electrostatic environment and in the realized trajectory, as well as slow drift and noise in magnetic fields and gradients. Such fluctuations produce \emph{stochastic} deviations of the spin Hamiltonian from its mean value. When averaged over many shuttling realizations, these stochastic contributions lead to a nonunitary, incoherent errors onto the qubit state. In this section we derive a compact circuit-level description of these incoherent effects in terms of an effective single-qubit Pauli channel, and we express the corresponding error probabilities in terms of experimentally meaningful fluctuation statistics.

\vspace{-1em}
\subsection{Noise Hamiltonian and random rotation angles}
\label{subsec:noise_angles}

We write the Hamiltonian during a shuttling segment of duration $\tau$ as
\begin{equation}
H(t)=H_{\rm det}(t)+\delta H(t),
\label{eq:H_det_plus_noise}
\end{equation}
where $H_{\rm det}(t)$ contains the deterministic evolution, $\delta H(t)=\frac{\hbar}{2}\,\delta\bm{\Omega}(t)\cdot\bm{\sigma}$ denotes the fluctuations. $\delta\bm{\Omega}(t)$ is the angular-frequency fluctuation vector representing the noisy spin precession induced by environmental variations. We assume $\langle\delta\bm{\Omega}(t)\rangle=\bm{0}$, so that $\delta H(t)$ represents zero-mean stochastic noise. The corresponding noise propagator is
\begin{equation}
U_{\rm n}=\mathcal{T}\exp\!\left[-\frac{i}{2}\int_0^\tau \delta\bm{\Omega}(t)\cdot\bm{\sigma}\,dt\right].
\label{eq:U_noise_timeordered}
\end{equation}
By approximating the noise as constant over the shuttling event, the time-ordering operator can be expressed as:
\begin{equation}
U_{\rm n}\approx \exp\!\left[-\frac{i}{2}\sum_{i\in\{x,y,z\}}\theta_i\,\sigma_i\right].
\label{eq:U_noise_angles}
\end{equation}
The random variables $\theta_i\equiv \int_0^\tau \delta\Omega_i(t)\,dt$ encode the net stochastic rotation accumulated during the shuttling event.

\subsection{Pauli-channel mapping}
\label{subsec:pauli_channel}

To connect the noise angles $\theta_i$ to a circuit-level error model, we now evaluate their effect on the qubit state. For a given realization, the state density matrix $\rho$ transforms as $\rho\mapsto U_{\rm n}\rho U_{\rm n}^\dagger$. Expanding to second order in the small angles $\theta_i$ yields
\begin{equation}
\rho' \approx \rho - i[A,\rho] + A\rho A -\frac{1}{2}\{A^2,\rho\},
\label{eq:rho_prime_second_order}
\end{equation}
where $A\equiv \frac{1}{2}\sum_i \theta_i\sigma_i$, up to $\mathcal{O}(\theta^3)$. Using $\sigma_i\sigma_j=\delta_{ij}\mathbb{I}+i\sum_k\varepsilon_{ijk}\sigma_k$ and the symmetry of $\theta_i\theta_j$, one finds $A^2=\frac{1}{4}(\sum_i\theta_i^2)\mathbb{I}$. Consequently,
\begin{equation}
\rho' \approx \rho - i[A,\rho]
+\frac{1}{4}\sum_{i,j}\theta_i\theta_j\,\sigma_i\rho\sigma_j
-\frac{1}{4}\left(\sum_i\theta_i^2\right)\rho.
\label{eq:rho_prime_explicit}
\end{equation}

To determine the effective evolution one must compute the average $\langle\rho'\rangle$ over these noise realizations. Assuming zero-mean noise fluctuations, $\langle\theta_i\rangle=0$, the first-order commutator term vanishes upon averaging. The remaining effect of the noise is given by the covariance matrix $C_{ij}\equiv \langle \theta_i\theta_j\rangle$. Finally, the mapping becomes:
\begin{equation}
\mathcal{E}_{\rm inc}(\rho)\equiv \langle\rho'\rangle
\approx
\rho+\frac{1}{4}\sum_{i,j}C_{ij}\,\sigma_i\rho\sigma_j
-\frac{1}{4}\left(\sum_i C_{ii}\right)\rho.
\label{eq:general_inc_map}
\end{equation}
In the Pauli-twirled approximation, one assumes that errors along different Pauli axes are uncorrelated, meaning $C_{ij}\approx 0$ for $i\neq j$. We note that while this is a standard simplifying assumption, it may not strictly apply during shuttling operations, where the coherent evolution and interplay of magnetic gradients and spin-orbit coupling can induce correlations between different noise components. Nevertheless, under this assumption, Eq.~\eqref{eq:general_inc_map} reduces to a Pauli channel
\begin{equation}
\mathcal{E}_{\rm P}(\rho)=
(1-p_X-p_Y-p_Z)\rho+p_XX\rho X+p_YY\rho Y+p_ZZ\rho Z,
\label{eq:Pauli_channel}
\end{equation}
with error probabilities
\begin{equation}
p_i=\frac{1}{4}\,C_{ii}=\frac{1}{4}\langle \theta_i^2\rangle.
\label{eq:pi_from_cov}
\end{equation}
The term $C_{ii}$ represents the variance of the stochastic rotation angle $\theta_i$ acquired during the shuttling event. Physically, this term encapsulates the effect of all noise sources such as fluctuating magnetic fields, uncertain electron trajectories through MM gradients, and variable SOC fields. An increase in this variance directly increases the corresponding Pauli error probability $p_i$.

\subsection{Trajectory-induced incoherent noise}
\label{subsec:traj_inc}

Trajectory deviations convert spatial inhomogeneity of the micromagnet field into stochastic spin rotations. Writing $\mathbf{r}(t)=\mathbf{r}_0(t)+\delta\mathbf{r}(t)$ and linearizing the MM field, using the gradient tensor $b_{ij}\equiv \partial B_i/\partial r_j$,
\begin{equation}
B_i(\mathbf{r}(t))\simeq B_i(\mathbf{r}_0(t))+\sum_{j} b_{ij}(\mathbf{r}_0(t))\,\delta r_j(t),
\label{eq:B_linear_traj}
\end{equation}
the corresponding fluctuation in the local Larmor precession frequency, with $\gamma\equiv g\mu_B/\hbar$, is
\begin{equation}
\delta\Omega_i^{(\delta\hat{x})}(t)=\gamma \sum_j b_{ij}(\mathbf{r}_0(t))\,\delta r_j(t).
\label{eq:dOmega_traj}
\end{equation}
The integrated stochastic angles are therefore
\begin{equation}
\theta_i^{(\delta\hat{x})}
=\int_0^\tau \delta\Omega_i^{(\delta\hat{x})}(t)\,dt
=\gamma\sum_j\int_0^\tau b_{ij}(\mathbf{r}_0(t))\,\delta r_j(t)\,dt.
\label{eq:theta_traj}
\end{equation}

In the quasi-static limit, one may approximate $b_{ij}(\mathbf{r}_0(t))\approx b_{ij}$ and $\delta r_j(t)\approx \delta r_j$, giving $\theta_i^{(\delta\hat{x})}\approx \gamma\tau\sum_j b_{ij}\delta r_j$. If the components $\delta r_j$ are uncorrelated, then
\begin{equation}
C_{ii}^{(\delta\hat{x})}\approx \gamma^2\tau^2\sum_j b_{ij}^2\,\langle \delta r_j^2\rangle,
\end{equation}
Using this variance, the corresponding diagonal Pauli error probability for the $i^{\text{th}}$ axis in the quasi-static limit simplifies to:
\begin{equation}
p_i^{(\delta\hat{x})}\approx \frac{1}{4}C_{ii}^{(\delta\hat{x})}.
\label{eq:pi_traj_qs}
\end{equation}

When the external magnetic field is applied along the $y$-axis, it defines the quantization axis. In this reference frame, phase errors ($Z$-type) correspond to fluctuations in the longitudinal field component ($B_y$), while bit-flip errors ($X$ and $Y$-type) correspond to fluctuations in the transverse components (e.g., $B_x$ and $B_z$). Assuming that the dominant micromagnet gradients are $b_{yx}$, $b_{yz}$, $b_{xy}$, and $b_{zy}$, and defining the variance of the trajectory fluctuations as $\langle\delta r_j^2\rangle \equiv (\delta j)^2$. If $b_{yy}$ is negligible but we still include the effect of $b_{xy}$ on the phase error rate (for instance, via a second-order coupling mechanism not explicit in the linear gradient expansion), the empirical bias form for trajectory noise coupling with field gradients becomes:
\begin{equation}
\eta_{\delta\hat{x}} = \frac{p_Z}{p_X} \approx \frac{(b_{yx}\delta x)^2 + (b_{yz}\delta z)^2 + (b_{xy}\delta y)^2}{(b_{zy}\delta y)^2 + (b_{xy}\delta y)^2}.
\label{eq:eta_bias}
\end{equation}
The equation above reveals the deep connection between spatial confinement and noise structure. If the quantum dot suffers predominantly from transverse stochastic motion ($\langle\delta x^2\rangle, \langle\delta z^2\rangle \gg \langle\delta y^2\rangle$), the numerator scales up rapidly, heavily favoring phase errors ($Z$-type) and enforcing a highly biased noise regime ($\eta_{\delta\hat{x}} \gg 1$). Conversely, if longitudinal fluctuations along the channel dominate ($\langle\delta y^2\rangle \gg \langle\delta x^2\rangle, \langle\delta z^2\rangle$), the bias limits to $\eta_{\delta\hat{x}} \approx (b_{xy}/b_{zy})^2$. This specific noise asymmetry, where trajectory noise couples strongly to field gradients, typically results in a biased noise channel for which non-CSS codes are particularly well suited.

\subsection{Field-induced incoherent noise}
\label{subsec:field_inc}

We consider the dominant source of magnetic-field fluctuations for a shuttling electron: drift or noise in the spatial gradient tensor itself. Waited-time fluctuations of the external magnetic field are typically slower and smaller in magnitude for these architectures and are thus neglected here. If the MM gradient coefficients fluctuate as $b_{ij}(t)=\overline{b}_{ij}(t)+\delta b_{ij}(t)$, then along the path the induced field fluctuation is
\begin{equation}
\delta B_i^{(\delta b)}(t)=\sum_j \delta b_{ij}(t)\,r_{0,j}(t),
\label{eq:dB_from_db}
\end{equation}
corresponding to an angular-frequency fluctuation $\delta\Omega_i^{(\delta b)}(t)=\gamma\,\delta B_i^{(\delta b)}(t)$.
The corresponding angles are
\begin{equation}
\theta_i^{(\delta b)}=\gamma\sum_j\int_0^\tau \delta b_{ij}(t)\,r_{0,j}(t)\,dt.
\label{eq:theta_db}
\end{equation}

In a quasi-static approximation with $r_{0,j}(t)\approx r_{0,j}$ and $\delta b_{ij}(t)\approx \delta b_{ij}$, one has $\theta_i^{(\delta b)}\approx \gamma\tau\sum_j r_{0,j}\delta b_{ij}$ and, for uncorrelated $\delta b_{ij}$,
\begin{equation}
C_{ii}^{(\delta b)}\approx \gamma^2\tau^2\sum_j r_{0,j}^2\,\langle (\delta b_{ij})^2\rangle.
\label{eq:pi_db_qs}
\end{equation}
This yields a Pauli error probability $p_i^{(\delta b)}\approx \frac{1}{4}C_{ii}^{(\delta b)}$.

Similar to the trajectory-induced noise case, when the external magnetic field is applied along the $y$-axis (defining the quantization axis), phase errors ($Z$-type) correspond to fluctuations in the longitudinal gradients, while bit-flip errors ($X$ and $Y$-type) correspond to fluctuations in transverse gradients. Let us define the relevant trajectory coordinate scales as $x_0, y_0, z_0$ and the variance of specific gradient components as $(\delta b_{ij})^2 \equiv \langle (\delta b_{ij})^2\rangle$. The resulting noise bias from gradient fluctuations, $\eta_{\delta\bar{b}}$, is given by the ratio of phase to bit-flip error probabilities:
\begin{equation}
\eta_{\delta\bar{b}} = \frac{p_Z}{p_X + p_Y} = \frac{x_0^2(\delta b_{yx})^2 + z_0^2(\delta b_{yz})^2}{y_0^2\big[(\delta b_{zy})^2 + (\delta b_{xy})^2\big]}.
\label{eq:eta_bias_db}
\end{equation}
This explicit formula similarly demonstrates how the characteristic dimensions of the electron wavepacket dictate the resulting error bias from background gradient drift. When the transverse spatial extent of the shuttling orbit is large compared to the longitudinal extent ($x_0^2, z_0^2 \gg y_0^2$), fluctuations in the longitudinal gradients strictly drive phase noise, leading to high asymmetry. Conversely, if the longitudinal extent dominates ($y_0^2 \gg x_0^2, z_0^2$), bit-flip errors become more prevalent, actively suppressing the bias. Adjusting the dynamic electrostatic confinement therefore provides a clear path towards tailoring the error bias.

\subsection{SOC-induced incoherent noise}
\label{subsec:soc_inc}

After subtracting any deterministic SOC contribution, SOC fluctuations enter as $\delta H_{\delta\hat{v}}(t)=\frac{\hbar}{2}\,\delta\bm{\Omega}_{\delta\hat{v}}(t)\cdot\bm{\sigma}$,
where $\langle\delta\bm{\Omega}_{\delta\hat{v}}(t)\rangle=\bm{0}$. To see explicitly how velocity fluctuations drive these errors, we substitute $p_i(t) = m^*(v_i(t) + \delta v_i(t))$ into the SOC Hamiltonian (Eq.~\ref{eq:H_SO}), yielding the fluctuation vector:
\begin{equation}
\begin{split}
\delta\bm{\Omega}_{\delta\hat{v}}(t) &= \frac{2m^*}{\hbar} \Big[ (\alpha_R \delta v_y(t) + \beta_D \delta v_x(t))\hat{x} \\
&\quad - (\alpha_R \delta v_x(t) + \beta_D \delta v_y(t))\hat{y} \Big].
\end{split}
\label{eq:dOmega_SO_explicit}
\end{equation}
The integrated angles are
\begin{equation}
\theta_i^{(\delta\hat{v})}\equiv \int_0^\tau \delta\Omega_{{\delta\hat{v}},i}(t)\,dt,
\label{eq:theta_SO}
\end{equation}
As before, we omit the off-diagonal components of the SOC fluctuation covariance, $C_{ij}^{(\delta\hat{v})} = \langle \theta_i^{(\delta\hat{v})}\theta_j^{(\delta\hat{v})} \rangle$, focusing on the diagonal basis where the Pauli error probability is
\begin{equation}
p_i^{(\delta\hat{v})}=\frac{1}{4}C_{ii}^{(\delta\hat{v})}.
\label{eq:pi_SO}
\end{equation}
As with the trajectory and field deviations, we can evaluate the resulting noise bias from SOC fluctuations, $\eta_{\delta\hat{v}}$. Using the Rashba and Dresselhaus strengths ($\alpha_R$ and $\beta_D$) and the corresponding fluctuation variances, the asymmetry in the noise is given by:
\begin{equation}
\eta_{\delta\hat{v}} \equiv \frac{p_Z}{p_Y} = \frac{\alpha_R^2\langle\delta x^2\rangle + \beta_D^2\langle\delta y^2\rangle}{\beta_D^2\langle\delta x^2\rangle + \alpha_R^2\langle\delta y^2\rangle}.
\label{eq:eta_bias_soc}
\end{equation}
This expression reveals how the noise asymmetry depends strongly on the physical direction of the velocity fluctuations. If the transport suffers predominantly from transverse stochastic motion ($\langle\delta x^2\rangle \gg \langle\delta y^2\rangle$), the bias converges to $\eta_{\delta\hat{v}} \approx (\alpha_R / \beta_D)^2$, dictated by the ratio of the Rashba to Dresselhaus coupling strengths. Conversely, if longitudinal velocity fluctuations along the shuttling channel dominate ($\langle\delta y^2\rangle \gg \langle\delta x^2\rangle$), the bias limits to $\eta_{\delta\hat{v}} \approx (\beta_D / \alpha_R)^2$. This pronounced geometric dependence offers an additional knob for noise tailoring. For instance, by aligning the shuttling channel appropriately with respect to the crystallographic axes and the external magnetic field, one can systematically tailor the SOC-induced bias to better match the capabilities of a specific topological code.

\subsection{Composite incoherent channel for a shuttling segment}
\label{subsec:composite_inc}

When the stochastic contributions from trajectory deviations, gradient fluctuations, and SOC are independent, the net angles add,
\begin{equation}
\theta_i=\theta_i^{(\delta\hat{x})}+\theta_i^{(\delta b)}+\theta_i^{(\delta\hat{v})},
\end{equation}
and the covariance matrix is additive,
\begin{equation}
C_{ij}=C_{ij}^{(\delta\hat{x})}+C_{ij}^{(\delta b)}+C_{ij}^{(\delta\hat{v})}.
\label{eq:cov_additive}
\end{equation}
In the diagonal-covariance approximation, this implies additivity of Pauli error probabilities at second order,
\begin{equation}
p_i \approx p_i^{(\delta\hat{x})}+p_i^{(\delta b)}+p_i^{(\delta\hat{v})},
\qquad i\in\{x,y,z\}.
\label{eq:pi_additive}
\end{equation}
Equations~\eqref{eq:Pauli_channel} and \eqref{eq:pi_additive} define the effective incoherent single-qubit noise channel associated with one shuttling segment within the circuit-level model. Crucially, because this composite error model takes the form of a Pauli channel, the error probabilities $p_X, p_Y$, and $p_Z$ can in principle be estimated directly \emph{in situ} from the syndrome measurement statistics of the stabilizer code itself \cite{spitz2021pauli}. This capability is particularly advantageous for shuttling based architectures, as it allows online tracking and optimal decoder adaptation  without the need for offline tomographic calibration. An overall effective noise bias for the complete shuttling operation can simply be defined by the ratio of the total longitudinal to transverse error probabilities:
\begin{equation}
\eta_{\text{comp}} \equiv \frac{p_Z}{p_X + p_Y}.
\label{eq:eta_bias_comp}
\end{equation}

\section{IV. Qubit Mapping and Shuttling Protocol}
We provide a mapping of the data qubits and check qubits of a rotated surface code (RSC), to a $2 \times N$ `railway'.
The data and the checks are mapped to separate `rails', such that checks are on the middle rail and data on one of the side rails (see Fig: \ref{fig:2N}).
We assume that shuttling is possible only along a rail, while entangling is possible only across the rails, i.e., perpendicular to the shuttling direction.

The data and checks from the same logical qubit system being mapped to different rails is the only viable mapping with which syndrome extraction can be executed in the railway architecture with only shuttling and clifford gate operations.
When represented as a bipartite Tanner graph, data and check qubits belonging to the same connected component---corresponding to a single logical qubit---cannot be assigned to the same rail.
This follows from the logic that length of any path between two vertices belonging to the same set of vertices in the bipartite graph, is even.
In the linear array architecture, length of any path from a rail to itself is always even, whereas length of any path to adjacent rail is always odd.

% The third rail is an avenue to shuttle in data qubits of a different logical qubit system and execute logical operation between \bl{the two sets of data qubits} \st{logical qubits}.
% We do not explore it in this paper, however, it is an interesting work to explore in the future.
% \bl{(data and check in the same code cannot belong to the same rail}
% \bl{(proof? with tanner graph. not included in this paper).}

Syndrome extraction of the code is carried out by shuttling either the data or check as a block, along the rail according to the \textit{train schedule} shuttling protocol that we propose in this paper.
The protocol will include shuttling instances and entanglement instances, where one type of qubit (data or check, depending upon the configuration) is shuttled after a entanglement instance to the position of the next entanglement instance.

An \textit{overlap} is a state of two qubits belonging to different but adjacent rails, being in the same column on the railway.
In a given entanglement instance, entanglements are carried out between the data and their corresponding check qubits, if they \textit{overlap} across the rails in that instance.

\subsection{Rotated Surface code (RSC)}
\begin{figure}[h!]
    \centering
    \input{figures/XZZXd5}
    \vspace{.5em}
    \caption{
    $d=5$ XZZX rotated surface code is shown here. The light and dark gray shaded faces/plaquettes corresponds to the difference in behavior of logical operators with respect to them. 
    % This is crucial for the choice of syndrome extraction order for each type of plaquettes,to avoid hook errors aligning with the logical operators.
    }
    \label{fig:xzzxd5}
\end{figure}
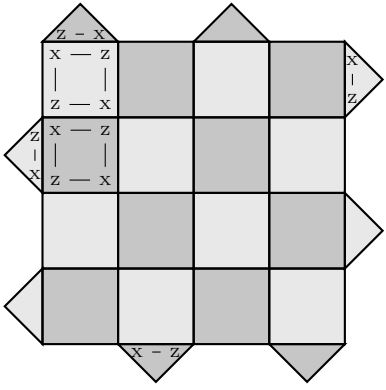

A rotated surface code (RSC) encodes a single logical qubit into an $n$-qubit system housed in a square lattice \cite{decoders,fowler2012surface}.
The stabilizers of the code correspond to each face (or \textit{plaquette}) of the lattice, with some of them acting on the edges of the lattice as well.
An XZZX RSC with distance $d=5$ is shown as a checkerboard plaquette lattice in Fig. \ref{fig:xzzxd5}.

 % Let $O_i$ be the unitary operator such that the single-qubit operator $O$ acts on the $i^{\text{th}}$ data qubit and the identity operator $I$ acts on the remaining qubits.
 % \begin{equation}
 % O_i=\bigotimes_{1,..,i-1} I \otimes O \otimes \bigotimes_{i+1,...,n} I
 % \end{equation}
The stabilizers corresponding to a plaquette $p$ (blue plaquettes in Fig. \ref{fig:hook_error}) and a plaquette $q$ (red plaquettes in Fig. \ref{fig:hook_error}) of a CSS rotated surface code $[[d^2, 1, d]]_2$ are defined as 
\begin{equation}
S_X^p=\prod_i X_i, \forall i \in P 
\end{equation}
\begin{equation}
S_Z^q=\prod_i Z_i, \forall i \in Q
\end{equation}
where $P$ is the set of data qubits on the corners of the plaquette $p$ (and likewise for $Q$).

The stabilizer corresponding to a plaquette of an XZZX rotated surface code (Fig: \ref{fig:xzzxd5}) is given by
\begin{equation}
S^p= X_iZ_jZ_kX_l 
\end{equation}
where $i,j,k,l$ are the qubits in north-west, north-east, south-west and south-east corner of the plaquette $p$, respectively.
For plaquettes with qubits only in \{$i$, $j$\}, \{$i$, $k$\}, \{$j$, $l$\} and \{$k$, $l$\}, the stabilizers are $X_iZ_j$, $X_iZ_k$, $Z_jX_l$ and $Z_kX_l$, respectively.
% For plaquettes with only two supported qubits, the 

For any two stabilizers $S^p, S^q \in S$, the set of stabilizers of a code, the commutation of $S^p$ and $S^q$ is given by \begin{equation}[S^p,S^q]=0\end{equation}
From Fig. \ref{fig:xzzxd5} and \ref{fig:hook_error}, it can be seen that the commutation constraints of both the RSCs are satisfied.
Stabilizers corresponding to plaquettes that meet at a corner share the same Pauli operator on the common data qubit.
Stabilizers corresponding to plaquettes that share an edge anti-commute on both common data qubits, ensuring the overall commutation between the stabilizers.
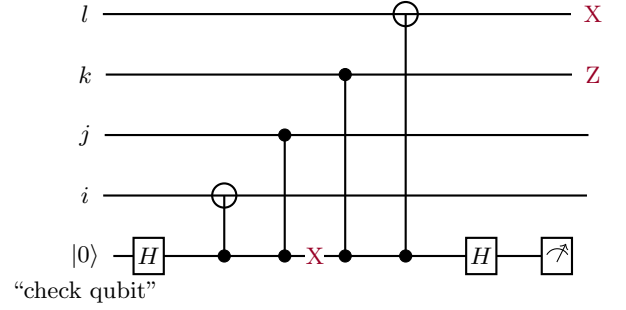
\begin{figure}[h!]
\input{figures/XZZXSyndromeExtractionCircuit}
    \vspace{-1em}
    \caption{
    Syndrome extraction circuit for an $X_i Z_j Z_k X_l$ stabilizer plaquette in the XZZX rotated surface code. 
    The check qubit is initialized in $\ket{+}$, accomplished by a Hadamard gate applied to $\ket{0}$. 
    It is sequentially entangled with the data qubits $i, j, k, l$ using CNOT gates (for $X$ operator of the stabilizer) and CZ gates (for $Z$ operator of the stabilizer), before being measured in the X-basis. 
    The red `X' illustrates a single physical fault occurring on the check qubit mid-cycle, specifically between the second and third entangling operations. 
    During the remainder of the circuit, this single fault propagates to the data qubits $k$ and $l$, manifesting as a correlated weight-2 data error ($Z_k X_l$). 
    These correlated errors resulting from a single circuit-level fault are known as hook errors.
    }
    \label{fig:HookErrorCircuit}
\end{figure}

Note that in the syndrome extraction circuit (shown in Fig. \ref{fig:HookErrorCircuit}), an $X$ error on the check qubit occurring at the midpoint of an extraction cycle (between the second and third entanglement) creates errors on two data qubits by the end of the cycle.
Such errors are called \textit{hook errors}, which, if aligned along the direction of the logical operators, lower the \textit{effective distance} of the code \cite{Dastbasteh2025generalized} (see Fig. \ref{fig:hook_error}).

\begin{figure}[h!]
    \centering
    \input{figures/HookErrord5}
    \vspace{.5em}
    \caption{Three single qubit errors: two hook errors and a single qubit error on a data qubit (shown as red X's), forming an undetectable logical error on a $d=5$ CSS rotated surface code. This choice syndrome extraction order aligns the hook error along the direction of the logical X operator.}
    \label{fig:hook_error}
\end{figure}
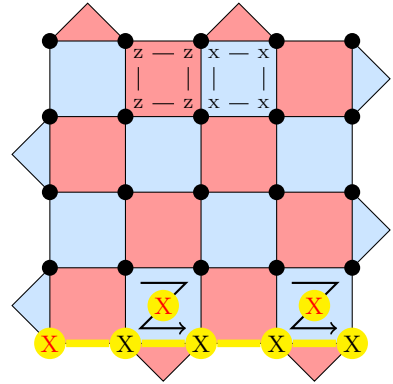

\begin{figure}[h!]%{0.9\textwidth}
    \centering
    \resizebox{\columnwidth}{!}{%
    \input{figures/XZZXplaquettes}}
    \vspace{.5em}
    \caption{The extraction order to avoid hook errors aligning with any logical operator for each plaquettes are shown here. The dark gray colored plaquettes and the light gray colored plaquettes have to have different set of extraction orders. The extraction orders in orange are suitable for all plaquettes.}
    \label{fig:XZZXplaquettes}
\end{figure}
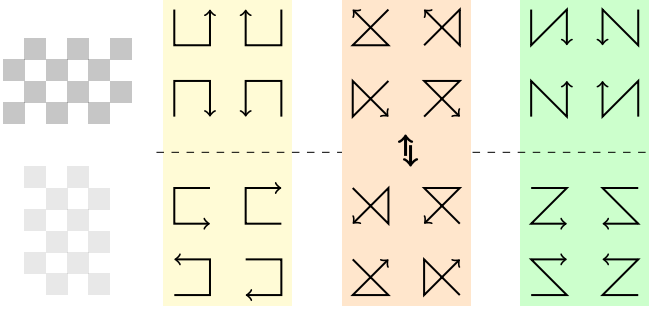

For the RSC, the syndrome extraction order that do not align hook errors along the logical operator direction is shown in Fig. \ref{fig:XZZXplaquettes}.
The extraction orders for the dark gray plaquettes in the XZZX code apply equally to the $Z$-stabilizers in the CSS RSC (and analogously for the light gray plaquettes and $X$-stabilizers).

\subsection{Shuttling Protocol: The Train Schedule}
We propose a shuttling protocol to perform syndrome extraction while satisfying the commutation constraints of the stabilizer code.
% In our current model of the \textit{Train Schedule}, we are shuttling all qubits of a type (either data or check) in only in one direction as a block in a round of syndrome extraction.
To minimize idling and shuttling overhead, we adopt a unidirectional `Train Schedule' that completes all required entanglements in a single pass while maintaining stabilizer commutation.
% We reason that there is no advantage in back shuttling (going back and forth) to do an entanglement in a round when the same can be achieved by doing the same at their first encounter/overlap, as long as all the commutation constraints can be satisfied by the protocol.
% Next we show that the commutation constraints of the stabilizer codes are satisfied in the \textit{Train Schedule.}

\begin{figure}[h!]
    \centering
    \input{figures/XZZXcommutetrain}
    \vspace{.5em}
    \caption{Syndrome extraction of stabilizers corresponding to check qubits: $q_1$ and $q_2$, in a data qubit shuttling configuration. For stabilizer corresponding to check qubit $q_1$, the desired extraction order of data qubits $i-k-j-l$ is achieved through their positional order of $i-k-j-l$ in the rail. Additionally, the commutation constraints are satisfied by the \textit{Train Schedule}, because the common data qubits on which both stabilizers corresponding to $q_1$ and $q_2$ has support on: $i$ and $j$, overlap with the data qubits $q_1$ and $q_2$ in the same order.}
    \label{fig:commutetrain}
\end{figure}
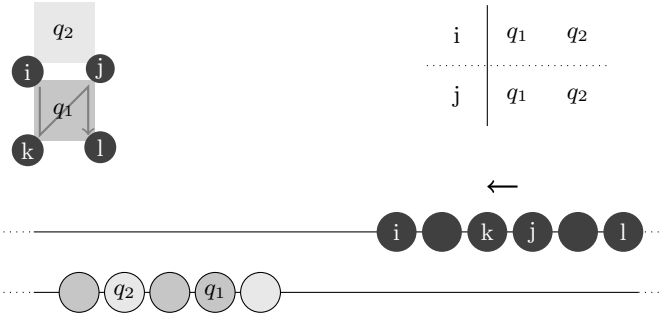

In syndrome extraction circuits, the commutation of two stabilizers of the RSC, sharing support on two data qubits, is satisfied as long as the entanglement of their corresponding check qubits with the shared data qubits occurs in the same order.
Within the \textit{Train schedule}, the shared data qubits interact with the corresponding check qubits in a consistent, identical sequence. This is because the Train schedule is unidirectional, and the entangling operations are carried out during the first instance of overlap. For any stabilizer code, the \textit{Train Schedule} ensures that commutation constraints are intrinsically satisfied.

In Fig. \ref{fig:commutetrain}, to implement an extraction order of $i-k-j-l$, we simply arrange the positional order of the data qubits in the rail as $i-k-j-l$ under the \textit{Train Schedule} protocol.
This is because in the protocol, the entanglements are performed at the first instance of the corresponding data-check qubit overlap.
Thereby, the check qubit $q_1$ entangles with data qubits in the order: $i-k-j-l$.
Generalizing this for all plaquettes, we see that the extraction order of plaquettes can be converted into a positional order rule for the data qubits in the rail.
However, the rules can be contradictory and their corresponding plaquette pattern pairs invalid, as is the case with the choice of extraction orders (which we call \textit{plaquette patterns}) in yellow in Fig. \ref{fig:XZZXplaquettes}.
The existence of loops of the positional order rules means that the qubits cannot be arranged in 1-dimensional order satisfying them.
In the figure, the \textit{plaquette patterns} in orange are a viable option, but they do not provide a single and convenient order that can be implemented in our mapping.
The plaquette patterns pairs in green which have the same starting and ending points are valid pairs and give non-contradictory positional rules.

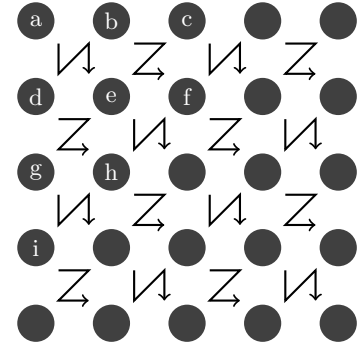
\begin{figure}[h!]
    \centering
    \input{figures/plaquetted5}
    \vspace{.5em}
    \caption{RSC code with syndrome extraction orders chosen from a pair of plaquette patterns in the green section of Fig: \ref{fig:XZZXplaquettes}.}
    \label{fig:plaquetted5}
\end{figure}
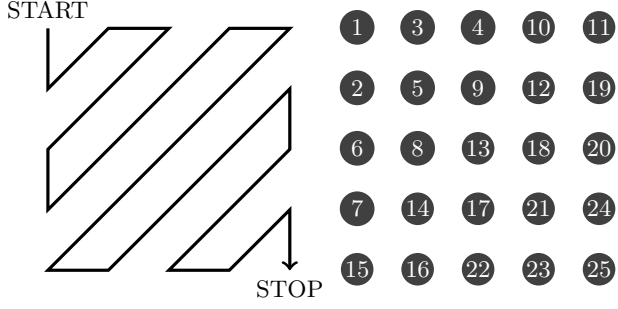
\begin{figure}[h!]
    \centering
    \input{figures/Snake_order}
    \input{figures/Snake_indices}
    \vspace{.5em}
    \caption{Data qubit order satisfying the order rules in Fig: \ref{fig:plaquetted5}. The \textit{snake order} is a general ordering of RSC data qubits for any distance code.}
    \label{fig:datasnake}
\end{figure}

Fig: \ref{fig:plaquetted5} shows an example plaquette pattern pair from the green section of Fig: \ref{fig:XZZXplaquettes}, implemented in a $d=5$ RSC.
The positional mapping is determined through an iterative assignment: we initialize by assigning the first position to qubit a and recursively determine subsequent positions based on the extraction order constraints.
% The final position in the rail for data qubits, can be obtained by solving for the positional order rules iteratively.
To start, except qubit $a$, all other qubits have to follow at least one other qubit.
We assign position $1$ to qubit $a$ and proceed with our reasoning.
Excluding qubit $a$ (it's final position has been assigned and we can take it out of our consideration), except for qubit $d$, all other qubits have to follow at least one other qubit.
So we can assign position $2$ to qubit $d$.
Following this logic iteratively, we can assign the next positions in the order $b$, $c$, $e$, $g$, $i$, $h$, $f$ and so on.
The resulting positional order for $d=5$ RSC is shown in Fig: \ref{fig:datasnake}
% Imposing the \textit{Train Schedule} on to our syndrome extraction order we studied 

The same order, which we term the \textit{snake order}, can be extended to any distance RSC.
Using a different choice of plaquette pattern pairs in the green section of Fig: \ref{fig:XZZXplaquettes}, we can alter the snake order to start from any corner of the square lattice and end in the diagonally opposite corner. 

\begin{figure}[h!]
    \centering
    \input{figures/XZZXallqubitorder}
    \vspace{.5em}
    \caption{Data and check qubit order (\textit{snake order} and \textit{mirrored snake order}) minimizing the shuttling instances and satisfying data qubit orders for the XZZX code. The same orders can be used for the CSS RSC and can also be intuitively extrapolated to any distance RSC.}
    \label{fig:xzzxallqubitorder}
\end{figure}
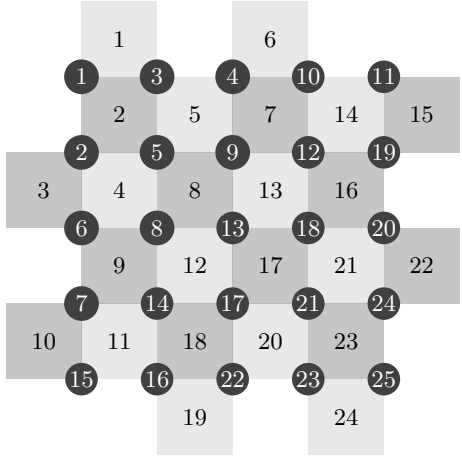

To minimize the shuttling instances, we choose the positional order of check qubits as shown in Fig: \ref{fig:xzzxallqubitorder}.
Note that the check qubit $1$ has to overlap data qubits $1$ and $3$, while check qubit $2$ has to overlap data qubits $1$ and $5$ (among others).
The positional order of check qubits are assigned in a \textit{mirrored snake order}, corresponding to the range of data qubit positions they need to overlap with.

\begin{figure}[htbp]
\resizebox{\linewidth}{!}{\input{figures/XZZXschedulegraph.tex}}
    \vspace{-1.5em}
    
  % \captionsetup{width=\linewidth}
    \caption{
    Map of Syndrome extraction for d=5 RSC in Train schedule, indexed as per Fig. \ref{fig:xzzxallqubitorder}.
    It shows the check qubits (horizontal axis, dark and light gray) and their corresponding data qubits (vertical axis, black) for syndrome extraction. 
    Each green dot is a CNOT or CZ gate, acting between a check qubit (horizontal axis) and the corresponding data qubit (vertical axis). \\
    \textbf{How to read the graph:} 
    The parameters $t$ and $x$ are illustratory and corresponds to the time and position of the shuttled check qubits.
    % The entanglement time is not considered here.
    The label `$x=0$' (`$t=4,x=0$') is the relative position of check qubits w.r.t data qubits, when check qubit $1-24$ exactly aligns with data qubits $1-24$.
    Each $45^{\circ} $ line corresponds to an entanglement instance. 
    The green dots incident on a line are the entanglements that are done in the corresponding entanglement instance.
    In a round of syndrome extraction, the check qubits are in a relative position `$x=4$' while `$t=0$', and proceed to `$x=-5$' at `$t=9$'.
    The reverse movement is also a valid shuttling protocol.
    % $x=0$ is the position of data qubits where qubit $1$ of data qubits overlaps with qubit $1$ of check qubits (it holds true for data qubits: $1-24$ with their corresponding check qubits: $1-24$). 
    }
    \label{fig:XZZXschedulegraph}
\end{figure}
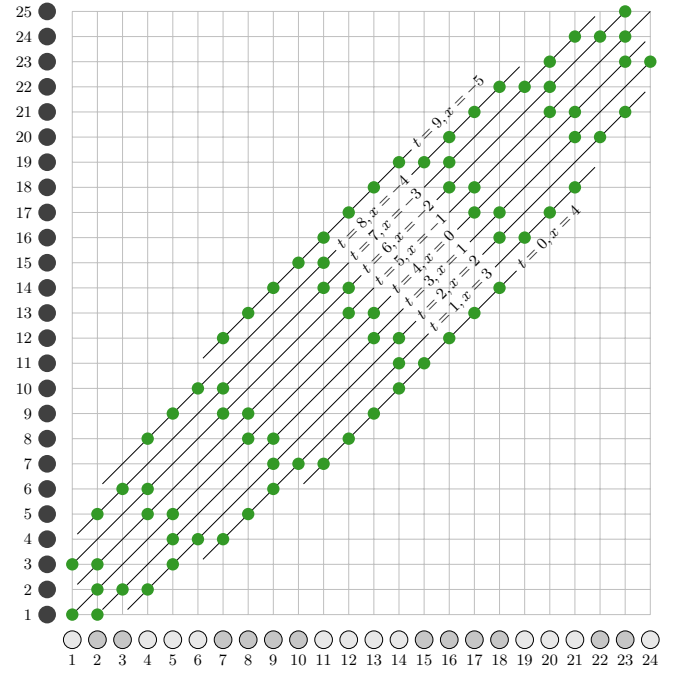

Fig: \ref{fig:XZZXschedulegraph} shows a graph with complete list of entanglements between the check and data qubits, in a single round of syndrome extraction.
In RSC, entanglement instances are all adjacent to each other, requiring only one shuttle of the data (or check) qubit block between them.
For illustration in the figure, a unit time ($t$) is taken as time to do one shuttling and the entanglements assumed are instantaneous, whereas the length of one shuttle is taken as a unit distance ($x$).
The $t$ and $x$ values illustrate where the data (or check) block will be with respect to the check (or data) qubit block.
To formally evaluate these operational sequences, we consider multiple architectural configurations: holding the data qubits stationary while shuttling the check layer versus moving the data over a static check layer. The ensuing circuit-level noise analysis critically depends on the idling logic bias ($\eta$) as well as the shuttling-induced error bias ($\eta_s$, which can be either X- or Z-biased), as we will explore in detail.
For a $d$ distance RSC, our mapping and shuttling protocols gives $2d$ number of entanglement instances and $2d-1$ number of shuttling instances.
The latter is easily understood as the $2d-1$ number of $1$ shuttles in between the $2d$ entanglement instances.
Additionally, there are $2d-2$ number of entanglements happening in every entanglement instance.
The shuttling direction is reversible for the given qubit orders, giving a similar syndrome extraction process.
As depicted in the graph of Fig. \ref{fig:XZZXschedulegraph}, we define a single shuttling instance as the time required to transport a qubit by the distance of one adjacent QD spacing. For rotated surface codes specifically, every such shuttling step is immediately followed by a entanglement instance.
In our shuttling protocol, the next round of syndrome extraction do not need additional time steps for shuttling the blocks to the initial position.
We can start the next round of syndrome extraction from the final position of the last round and with the shuttling direction reversed. It is worth noting that the subsequent direction-reversed syndrome extraction round is entirely analogous to the forward round, retaining the same protective properties regarding hook error mitigation, strict stabilizer commutation, and the minimization of total shuttling instances.

\begin{table*}[htbp]
\centering
\caption{Threshold values ($p_{th}\times10^3$) of rotated surface codes (RSC) under different configurations. 
$p_{th}$ corresponds to the physical error rate due to a two qubit gate operation on the physical qubits. 
For each configuration, threshold values are given for the two complementary memory directions - V (Vertical) and H (Horizontal) for the XZZX RSC, and Z (memory) and X (memory) for the CSS RSC. 
The table compares the standard Calderbank-Shor-Steane (CSS) surface code against the XZZX surface code variant. 
The operative noise parameters include the idling error and CZ bias ($\eta$) and the shuttling-induced error bias ($\eta_s$), which can be specifically directed, such as being X-biased ($\eta_X$) or Z-biased ($\eta_Z$).
% Threshold values ($p_{th}\times10^3$) of rotated surface codes (RSC) under different configurations. For each configuration, threshold values are given for the two complementary memory directions, V (Vertical) and H (Horizontal). For the standard Calderbank-Shor-Steane (CSS) code, these explicitly correspond to the X and Z logical memories, respectively. The table compares the CSS surface code against the XZZX surface code variant. The operative noise parameters include the idling error and CZ bias ($\eta$) and the shuttling-induced error bias ($\eta_s$), which can be specifically directed, such as being X-biased ($\eta_X$) or Z-biased ($\eta_Z$).
}
\label{tab:combined_thresholds}
\vspace{0.25cm}
\setlength{\tabcolsep}{6pt}
\small
\setCustomStrut
\begin{tabular}{|c|c|c|c|@{}c@{}|@{}c@{}|@{}c@{}|@{}c@{}|@{}c@{}|@{}c@{}|}
\hline
\multirow{2}{*}{\makebox[1.6cm][c]{Code}} & \multirow{2}{*}{\makebox[1.6cm][c]{Shuttled}} & \multirow{2}{*}{\makebox[1.6cm][c]{Bias Dir.}} & \multirow{2}{*}{\makebox[1.6cm][c]{$\eta$}} & \multicolumn{2}{c|}{$\eta_{s} = 0.5$} & \multicolumn{2}{c|}{$\eta_{s} = 10$} & \multicolumn{2}{c|}{$\eta_{s} = 100$} \\ \cline{5-10}
& & & & \makebox[1.4cm][c]{V/Z}& \makebox[1.4cm][c]{H/X}& \makebox[1.4cm][c]{V/Z}& \makebox[1.4cm][c]{H/X}& \makebox[1.4cm][c]{V/Z}& \makebox[1.4cm][c]{H/X} \\ \hline
\multirow{3}{*}{CSS} & \multirow{3}{*}{Data} & \multirow{3}{*}{X ($\eta_X$)} 
& 0.5 & \cellcolor{soothinggreen!7!soothingred}2.2 & \cellcolor{soothinggreen!7!soothingred}2.2 & \cellcolor{soothingred}$<2$ & \cellcolor{soothingred}3 & \cellcolor{soothingred}$<2$ & \cellcolor{soothingred}3 \\ \cline{4-10}
& & & 10 & \cellcolor{soothingred} \makebox[1.4cm][c]{3.2}& \cellcolor{soothingred} \makebox[1.4cm][c]{2}& \cellcolor{soothinggreen!27!soothingred} \makebox[1.4cm][c]{2.55}& \cellcolor{soothinggreen!27!soothingred} \makebox[1.4cm][c]{2.55}& \cellcolor{soothinggreen!30!soothingred} \makebox[1.4cm][c]{2.6}& \cellcolor{soothinggreen!30!soothingred} \makebox[1.4cm][c]{2.85} \\ \cline{4-10} 
& & & 100 & \cellcolor{soothinggreen!5!soothingred} \makebox[1.4cm][c]{3.4}& \cellcolor{soothinggreen!5!soothingred} \makebox[1.4cm][c]{2.1}& \cellcolor{soothinggreen!25!soothingred} \makebox[1.4cm][c]{2.8}& \cellcolor{soothinggreen!25!soothingred} \makebox[1.4cm][c]{2.5}& \cellcolor{soothinggreen!40!soothingred} \makebox[1.4cm][c]{2.8}& \cellcolor{soothinggreen!40!soothingred} \makebox[1.4cm][c]{2.85} \\ \hline 
\multirow{3}{*}{CSS} & \multirow{3}{*}{Data} & \multirow{3}{*}{Z ($\eta_Z$)} 
& 0.5 & \cellcolor{soothinggreen!7!soothingred}2.2 & \cellcolor{soothinggreen!7!soothingred}2.2 & \cellcolor{soothingred}$2.8$ & \cellcolor{soothingred}$<2$ & \cellcolor{soothingred}$3.1$ & \cellcolor{soothingred}$<2$ \\ \cline{4-10}
& & & 10 & \cellcolor{soothingred} \makebox[1.4cm][c]{$3$}& \cellcolor{soothingred} \makebox[1.4cm][c]{$2$}& \cellcolor{soothingred} \makebox[1.4cm][c]{$>5$}& \cellcolor{soothingred} \makebox[1.4cm][c]{$<2$}& \cellcolor{soothingred} \makebox[1.4cm][c]{$>5$}& \cellcolor{soothingred} \makebox[1.4cm][c]{$<2$} \\ \cline{4-10} 
& & & 100 & \cellcolor{soothinggreen!5!soothingred} \makebox[1.4cm][c]{$3.5$}& \cellcolor{soothinggreen!5!soothingred} \makebox[1.4cm][c]{$2.1$}& \cellcolor{soothingred} \makebox[1.4cm][c]{$>5$}& \cellcolor{soothingred} \makebox[1.4cm][c]{$<2$}& \cellcolor{soothingred} \makebox[1.4cm][c]{$>5$}& \cellcolor{soothingred} \makebox[1.4cm][c]{$<2$} \\ \hline 
\multirow{3}{*}{XZZX} & \multirow{3}{*}{Data} & \multirow{3}{*}{X ($\eta_X$)} 
& 0.5 & \cellcolor{soothinggreen!7!soothingred}2.2 & \cellcolor{soothinggreen!10!soothingred}2.3 & \cellcolor{soothinggreen!17!soothingred}2.5 & \cellcolor{soothinggreen!13!soothingred}2.4 & \cellcolor{soothinggreen!8!soothingred}2.25 & \cellcolor{soothinggreen!15!soothingred}2.45 \\ \cline{4-10}
& & & 10 & \cellcolor{soothinggreen!25!soothingred} \makebox[1.4cm][c]{2.5}& \cellcolor{soothinggreen!25!soothingred} \makebox[1.4cm][c]{2.5}& \cellcolor{soothinggreen!35!soothingred} \makebox[1.4cm][c]{2.7}& \cellcolor{soothinggreen!35!soothingred} \makebox[1.4cm][c]{2.75}& \cellcolor{soothinggreen!25!soothingred} \makebox[1.4cm][c]{2.5}& \cellcolor{soothinggreen!25!soothingred} \makebox[1.4cm][c]{2.75} \\ \cline{4-10} 
& & & 100 & \cellcolor{soothinggreen!30!soothingred} \makebox[1.4cm][c]{2.6}& \cellcolor{soothinggreen!30!soothingred} \makebox[1.4cm][c]{2.75}& \cellcolor{soothinggreen!35!soothingred} \makebox[1.4cm][c]{2.7}& \cellcolor{soothinggreen!35!soothingred} \makebox[1.4cm][c]{2.75}& \cellcolor{soothinggreen!35!soothingred} \makebox[1.4cm][c]{2.8}& \cellcolor{soothinggreen!35!soothingred} \makebox[1.4cm][c]{2.7} \\ \hline 
\multirow{3}{*}{XZZX} & \multirow{3}{*}{Data} & \multirow{3}{*}{Z ($\eta_Z$)} 
& 0.5 & \cellcolor{soothinggreen!7!soothingred}2.2 & \cellcolor{soothinggreen!7!soothingred}2.2 & \cellcolor{soothinggreen!13!soothingred}2.4 & \cellcolor{soothinggreen!22!soothingred}2.65 & \cellcolor{soothinggreen!23!soothingred}2.7 & \cellcolor{soothinggreen!20!soothingred}2.6 \\ \cline{4-10}
& & & 10 & \cellcolor{soothinggreen!15!soothingred} \makebox[1.4cm][c]{2.3}& \cellcolor{soothinggreen!15!soothingred} \makebox[1.4cm][c]{2.5}& \cellcolor{soothinggreen!60!soothingred} \makebox[1.4cm][c]{3.25}& \cellcolor{soothinggreen!60!soothingred} \makebox[1.4cm][c]{3.2}& \cellcolor{soothinggreen!70!soothingred} \makebox[1.4cm][c]{3.4}& \cellcolor{soothinggreen!70!soothingred} \makebox[1.4cm][c]{3.55} \\ \cline{4-10} 
& & & 100 & \cellcolor{soothinggreen!25!soothingred} \makebox[1.4cm][c]{2.6}& \cellcolor{soothinggreen!25!soothingred} \makebox[1.4cm][c]{2.5}& \cellcolor{soothinggreen!62!soothingred} \makebox[1.4cm][c]{3.4}& \cellcolor{soothinggreen!62!soothingred} \makebox[1.4cm][c]{3.25}& \cellcolor{soothinggreen!75!soothingred} \makebox[1.4cm][c]{3.6}& \cellcolor{soothinggreen!75!soothingred} \makebox[1.4cm][c]{3.5} \\ \hline 
\multirow{3}{*}{CSS} & \multirow{3}{*}{Check} & \multirow{3}{*}{X ($\eta_X$)} 
& 0.5 & \cellcolor{soothinggreen!22!soothingred}2.65 & \cellcolor{soothinggreen!23!soothingred}2.7 & \cellcolor{soothinggreen!20!soothingred}2.6 & \cellcolor{soothinggreen!18!soothingred}2.55 & \cellcolor{soothinggreen!23!soothingred}2.7 & \cellcolor{soothinggreen!25!soothingred}2.75 \\ \cline{4-10}
& & & 10 & \cellcolor{soothinggreen!23!soothingred} \makebox[1.4cm][c]{$>5$}& \cellcolor{soothinggreen!23!soothingred} \makebox[1.4cm][c]{2.45}& \cellcolor{soothinggreen!25!soothingred} \makebox[1.4cm][c]{$>5$}& \cellcolor{soothinggreen!25!soothingred} \makebox[1.4cm][c]{2.5}& \cellcolor{soothinggreen!25!soothingred} \makebox[1.4cm][c]{$>5$}& \cellcolor{soothinggreen!25!soothingred} \makebox[1.4cm][c]{2.5} \\ \cline{4-10} 
& & & 100 & \cellcolor{soothinggreen!23!soothingred} \makebox[1.4cm][c]{$>5$}& \cellcolor{soothinggreen!23!soothingred} \makebox[1.4cm][c]{2.45}& \cellcolor{soothinggreen!20!soothingred} \makebox[1.4cm][c]{$>5$}& \cellcolor{soothinggreen!20!soothingred} \makebox[1.4cm][c]{2.4}& \cellcolor{soothinggreen!23!soothingred} \makebox[1.4cm][c]{$>5$}& \cellcolor{soothinggreen!23!soothingred} \makebox[1.4cm][c]{2.45} \\ \hline 
\multirow{3}{*}{CSS} & \multirow{3}{*}{Check} & \multirow{3}{*}{Z ($\eta_Z$)} 
& 0.5 & \cellcolor{soothinggreen!25!soothingred}$2.75$ & \cellcolor{soothinggreen!18!soothingred}$2.55$ & \cellcolor{soothinggreen!27!soothingred}$2.8$ & \cellcolor{soothinggreen!28!soothingred}$2.85$ & \cellcolor{soothinggreen!28!soothingred}$2.85$ & \cellcolor{soothinggreen!23!soothingred}$2.7$ \\ \cline{4-10}
& & & 10 & \cellcolor{soothinggreen!23!soothingred} \makebox[1.4cm][c]{$>5$}& \cellcolor{soothinggreen!23!soothingred} \makebox[1.4cm][c]{2.45}& \cellcolor{soothinggreen!27!soothingred} \makebox[1.4cm][c]{$>5$}& \cellcolor{soothinggreen!27!soothingred} \makebox[1.4cm][c]{2.55}& \cellcolor{soothinggreen!23!soothingred} \makebox[1.4cm][c]{$>5$}& \cellcolor{soothinggreen!23!soothingred} \makebox[1.4cm][c]{2.45} \\ \cline{4-10} 
& & & 100 & \cellcolor{soothinggreen!23!soothingred} \makebox[1.4cm][c]{$>5$}& \cellcolor{soothinggreen!23!soothingred} \makebox[1.4cm][c]{2.45}& \cellcolor{soothinggreen!18!soothingred} \makebox[1.4cm][c]{$>5$}& \cellcolor{soothinggreen!18!soothingred} \makebox[1.4cm][c]{2.35}& \cellcolor{soothinggreen!23!soothingred} \makebox[1.4cm][c]{$>5$}& \cellcolor{soothinggreen!23!soothingred} \makebox[1.4cm][c]{2.45} \\ \hline 
\multirow{3}{*}{XZZX} & \multirow{3}{*}{Check} & \multirow{3}{*}{X ($\eta_X$)} 
& 0.5 & \cellcolor{soothinggreen!10!soothingred}2.3 & \cellcolor{soothinggreen!10!soothingred}2.3 & \cellcolor{soothinggreen!17!soothingred}2.5 & \cellcolor{soothinggreen!17!soothingred}2.5 & \cellcolor{soothinggreen!18!soothingred}2.55 & \cellcolor{soothinggreen!17!soothingred}2.5 \\ \cline{4-10}
& & & 10 & \cellcolor{soothinggreen!57!soothingred} \makebox[1.4cm][c]{3.15}& \cellcolor{soothinggreen!57!soothingred} \makebox[1.4cm][c]{3.2}& \cellcolor{soothinggreen!75!soothingred} \makebox[1.4cm][c]{3.5}& \cellcolor{soothinggreen!75!soothingred} \makebox[1.4cm][c]{3.5}& \cellcolor{soothinggreen!75!soothingred} \makebox[1.4cm][c]{3.5}& \cellcolor{soothinggreen!75!soothingred} \makebox[1.4cm][c]{3.5} \\ \cline{4-10} 
& & & 100 & \cellcolor{soothinggreen!82!soothingred} \makebox[1.4cm][c]{3.65}& \cellcolor{soothinggreen!82!soothingred} \makebox[1.4cm][c]{3.65}& \cellcolor{soothinggreen!85!soothingred} \makebox[1.4cm][c]{3.7}& \cellcolor{soothinggreen!85!soothingred} \makebox[1.4cm][c]{3.7}& \cellcolor{soothinggreen!90!soothingred} \makebox[1.4cm][c]{3.8}& \cellcolor{soothinggreen!90!soothingred} \makebox[1.4cm][c]{3.8} \\ \hline 
\multirow{3}{*}{XZZX} & \multirow{3}{*}{Check} & \multirow{3}{*}{Z ($\eta_Z$)} 
& 0.5 & \cellcolor{soothinggreen!10!soothingred}2.3 & \cellcolor{soothinggreen!10!soothingred}2.3 & \cellcolor{soothinggreen!17!soothingred}2.5 & \cellcolor{soothinggreen!17!soothingred}2.5 & \cellcolor{soothinggreen!17!soothingred}2.5 & \cellcolor{soothinggreen!17!soothingred}2.5 \\ \cline{4-10}
& & & 10 & \cellcolor{soothinggreen!50!soothingred} \makebox[1.4cm][c]{3}& \cellcolor{soothinggreen!50!soothingred} \makebox[1.4cm][c]{3.2}& \cellcolor{soothinggreen!73!soothingred} \makebox[1.4cm][c]{3.45}& \cellcolor{soothinggreen!73!soothingred} \makebox[1.4cm][c]{3.7}& \cellcolor{soothinggreen!75!soothingred} \makebox[1.4cm][c]{3.6}& \cellcolor{soothinggreen!75!soothingred} \makebox[1.4cm][c]{3.5} \\ \cline{4-10} 
& & & 100 & \cellcolor{soothinggreen!80!soothingred} \makebox[1.4cm][c]{3.65}& \cellcolor{soothinggreen!80!soothingred} \makebox[1.4cm][c]{3.6}& \cellcolor{soothinggreen!85!soothingred} \makebox[1.4cm][c]{4}& \cellcolor{soothinggreen!85!soothingred} \makebox[1.4cm][c]{3.7}& \cellcolor{soothinggreen} \makebox[1.4cm][c]{4}& \cellcolor{soothinggreen} \makebox[1.4cm][c]{4} \\ \hline 
\end{tabular}

\vspace{0.15cm}
\begin{minipage}{\linewidth}
\centering
\begin{tikzpicture}
\foreach \i in {0,1,...,50} {
  \pgfmathsetmacro{\pct}{\i*2}
  \fill[soothinggreen!\pct!soothingred] (\i*0.16, 0) rectangle (\i*0.16+0.17, 0.4);
}
\draw (0,0) rectangle (8.15,0 .4);
\node[anchor=north] at (0,-0.1) {\small $\leq 2$};
\node[anchor=north] at (4.075,-0.1) {\small Bottleneck Threshold $\min(V,H)$ ($p_{th} \times 10^3$)};
\node[anchor=north] at (8.15,-0.1) {\small $\geq 4$};
\end{tikzpicture}
\end{minipage}
\end{table*}
% \begin{table*}[htbp]
%     \centering
%     \caption{The qubit footprints and relative decrease in percentage with respect to the baseline symmetric $(\eta=0.5, \eta_Z=0.5)$ circuit-level noise model for various bias combinations, considered at physical error rates $p = 0.003$ and $p = 0.001$. \textcolor{red}{THIS TABLE IS UNCOLOURED FOR SOME REASON.}}
%     \label{tab:footprints_combined}
%     \vspace{0.25cm}
%     \renewcommand{\arraystretch}{1.2}
%     \begin{tabular}{cc|cccc|cccc}
% \hline\hline
%     \multirow{2}{*}{Bias Config} & \multirow{2}{*}{$(\eta, \eta_Z)$} & \multicolumn{4}{c|}{$p = 0.003$} & \multicolumn{4}{c}{$p = 0.001$} \\ \cline{3-10}
%     & & Megaquop & Dec. (\%) & Gigaquop & Dec. (\%) & Megaquop & Dec. (\%) & Gigaquop & Dec. (\%) \\
%     \hline
%     Symmetric& $(0.5,0.5)$  & $1799$& $\cdots$ & $5831$& $\cdots$ & $199$& $\cdots$ & $577$& $\cdots$ \\
%     Shuttling Bias& $(0.5,100)$  & $1567$& $\sim 13\%$ & $4999$& $\sim 14\%$ & $161$& $\sim 19\%$ & $511$& $\sim 11\%$ \\
%     Idling Bias& $(100,0.5)$  & $647$& $\sim 64\%$ & $2047$& $\sim 65\%$ & $127$& $\sim 36\%$ & $391$& $\sim 32\%$ \\
%     Full Bias& $(100,100)$  & $511$& $\sim 72\%$ & $1567$& $\sim 73\%$ & $97$& $\sim 51\%$ & $337$& $\sim 42\%$ \\
%      \hline\hline
% \end{tabular}
% \end{table*}

\begin{table*}[htbp]
    \centering
    \caption{The qubit footprints and relative decrease in percentage with respect to the baseline symmetric $(\eta=0.5, \eta_Z=0.5)$ circuit-level noise model for various bias combinations, considered at physical error rates $p = 0.003$ and $p = 0.001$.}
    \label{tab:footprints_combined}
    \vspace{0.25cm}
    \renewcommand{\arraystretch}{1.2}
    \begin{tabular}{cc|cccc|cccc}
\hline\hline
    \multirow{2}{*}{Bias Config} & \multirow{2}{*}{$(\eta, \eta_Z)$} & \multicolumn{4}{c|}{$p = 0.003$} & \multicolumn{4}{c}{$p = 0.001$} \\ \cline{3-10}
    & & Megaquop & Dec. (\%) & Gigaquop & Dec. (\%) & Megaquop & Dec. (\%) & Gigaquop & Dec. (\%) \\
    \hline
    Symmetric& $(0.5,0.5)$  & \cellcolor{soothingred}$1799$& \cellcolor{soothingred}$\cdots$ & \cellcolor{soothingred}$5831$& \cellcolor{soothingred}$\cdots$ & \cellcolor{soothingred}$199$& \cellcolor{soothingred}$\cdots$ & \cellcolor{soothingred}$577$& \cellcolor{soothingred}$\cdots$ \\
    Shuttling Bias& $(0.5,100)$  & \cellcolor{soothinggreen!13!soothingred}$1567$& \cellcolor{soothinggreen!13!soothingred}$\sim 13\%$ & \cellcolor{soothinggreen!14!soothingred}$4999$& \cellcolor{soothinggreen!14!soothingred}$\sim 14\%$ & \cellcolor{soothinggreen!19!soothingred}$161$& \cellcolor{soothinggreen!19!soothingred}$\sim 19\%$ & \cellcolor{soothinggreen!11!soothingred}$511$& \cellcolor{soothinggreen!11!soothingred}$\sim 11\%$ \\
    Idling Bias& $(100,0.5)$  & \cellcolor{soothinggreen!64!soothingred}$647$& \cellcolor{soothinggreen!64!soothingred}$\sim 64\%$ & \cellcolor{soothinggreen!65!soothingred}$2047$& \cellcolor{soothinggreen!65!soothingred}$\sim 65\%$ & \cellcolor{soothinggreen!36!soothingred}$127$& \cellcolor{soothinggreen!36!soothingred}$\sim 36\%$ & \cellcolor{soothinggreen!32!soothingred}$391$& \cellcolor{soothinggreen!32!soothingred}$\sim 32\%$ \\
    Full Bias& $(100,100)$  & \cellcolor{soothinggreen!72!soothingred}$511$& \cellcolor{soothinggreen!72!soothingred}$\sim 72\%$ & \cellcolor{soothinggreen!73!soothingred}$1567$& \cellcolor{soothinggreen!73!soothingred}$\sim 73\%$ & \cellcolor{soothinggreen!51!soothingred}$97$& \cellcolor{soothinggreen!51!soothingred}$\sim 51\%$ & \cellcolor{soothinggreen!42!soothingred}$337$& \cellcolor{soothinggreen!42!soothingred}$\sim 42\%$ \\
     \hline\hline
\end{tabular}
\end{table*}

The substantial footprint reductions observed under biased noise configurations underscore the critical importance of hardware-aware code design. By strategically aligning the XZZX code's error-suppression capabilities with the dominant noise characteristics of the shuttling operations, one can dramatically decrease the physical resource overhead required for fault-tolerant quantum computation. This architectural co-design approach ensures that the railway structure operates at its maximum efficiency. To visualize these performance trends further, we present the detailed relationships between various noise biases and their corresponding logical error outcomes in the following section.

\section{V. Simulation with circuit level noise: Results}
The odd code distances $3$ to $17$ RSC are studied under check qubit shuttling and data qubit shuttling configurations.
Additionally, the shuttle noise is considered for both X and Z biases.
We have implemented the noisy extraction circuits to perform the sampling of the errors using the STIM software package \cite{Gidney2021stimfaststabilizer}, which considers the check measurements upon a set of syndrome extractions together with a final measurement of the data qubits. We do $3\times d$ number of rounds for our threshold estimates, with a minimum of $1$ million shots or $1$ thousand errors (logical), whichever is reached first.

\subsection{Noise Model}
We consider an hybrid biased-depolarizing circuit-level noise model appropriate for two-level system qubits in our simulations for the XZZX and CSS RSC \cite{Etxezarreta2026leveraging}. We decode the syndrome data by means of the \textit{pymatching} implementation of the minimum-weight perfect matching decoder \cite{decoders,pymatching}.
Under different configurations, we either shuttle the check qubits or the data qubits, and compare their performances.
Different configurations of bias has also been studied.

\begin{table}[htbp]%[hbt!]%
    \centering
    \caption{Tailored circuit-level noise model for the proposed
architecture based on silicon spin qubits}
    \label{tab:noisemodel}
    \vspace{0.25cm}
    \resizebox{\columnwidth}{!}{%
    \begin{tabular}{c|cc}
    \hline
        Error type & Error& Bias \\
        \hline
        Single qubit gate  & $\frac{p}{10}$& Depolarizing\\
         CZ gate&   $p$ & $\eta$ (Z bias)\\
         CNOT gate &   $p$ & Depolarizing\\
         Idling  & $\frac{p}{10}\times IF$& $\eta$ (Z-bias)\\
        Shuttling     & $\frac{p}{10}$&$\eta_X$ (X bias) or $\eta_Z$ (Z bias) \\
         Measurement &   $2p$& Bit flip error\\
         Initialization & $2p$ & Bit/Phase flip error\\
         \hline
    \end{tabular}}
\end{table}

% \textcolor{red}{I THINK THAT HERE SOMETHIN IS MISSING, PROBABLY TABLE IV, BUT I AM UNSURE}

Table \ref{tab:noisemodel} details the type of noises and the biases we have considered in our model. The parameter, $p$, is the probability of error (PoE) due to application of a two qubit gate and $IF$ is the Idling Factor (IF) and is given by
\begin{equation}IF = \frac{\text{PoE (the awaited event)}}{\text{PoE (two qubit gate operation)}}\end{equation}
This logic fundamentally assumes that in the low-error regimes considered ($p \ll 1$), error probabilities scale approximately linearly with the time spent executing a physical operation. Thus, normalizing the awaited event's error probability relative to that of a two-qubit gate effectively makes the IF a weight for the baseline idling error, proportionate to the relative duration of the awaited physical operation.
The idling noise is always Z biased, while shuttling has been studied under both X and Z biases.
CZ gates have been modeled as bias preserving. Idling and CZ bias ($\eta$) over a $100$ has a negligible effect on the threshold, as shown by \cite{Etxezarreta2026leveraging}. 
% We were able to confirm the same in our simulations. 
% The shuttling error has been studied under both $X$ and $Z$ biases. 
We have observed a similar saturation of threshold for all biases over $100$.

To ground our theoretical noise models in current experimental observations, Table \ref{tab:spinhex_noise} summarizes recent fidelity benchmarks across different semiconductor spin-qubit architectures. This data highlights the specific error rates and biases characteristic of different material platforms and operational modes.

\begin{table}[htbp]
    \centering
    \caption{Noise model for various spin-qubit architectures.}
    \label{tab:spinhex_noise}
    \vspace{0.25cm}
    \resizebox{\columnwidth}{!}{%
    \begin{tabular}{|c|c|c|c|c|}
    \hline
    Operation & SiMOS(n) & Si/SiGe(n) & Ge/SiGe(p) & SiMOS(p) \\
    \hline
    Init & $> 99\%$ \cite{Huang_spinHex} & -- & -- & -- \\
    \hline
    H & $99.9\%$ \cite{Stuyck_spinHex} & $99.7\%$ \cite{Philips_spinHex} & $99.97\%$ \cite{Wang_spinHex} & $99.5\%$ \cite{Bassi_spinHex} \\
    \hline
    CZ/CX & $> 98.4\%$ \cite{Tanttu_spinHex} & $> 99\%$ \cite{Xue_spinHex} & $99.3\%$ \cite{Wang_spinHex} & -- \cite{Geyer_spinHex} \\

    \hline
    Read & $99.2\%$ \cite{Oakes_spinHex} & $99\%$ \cite{Takeda_spinHex} & $90\%$ \cite{vanRiggelen_spinHex} & low \cite{Bassi_spinHex} \\
    \hline
    $T_1$ & $1 - 10^3\text{ ms}$ & $1 - 10^3\text{ ms}$ & $>1\text{ms}$ \cite{Hendrickx_spinHex} & $>10\mu\text{s}$ \cite{Camenzind_spinHex} \\
    \hline
    $T_2^*$ & $10 - 10^2\mu\text{s}$ & $10 - 10^2\mu\text{s}$ & $<400\text{ns}$ \cite{Hendrickx_spinHex} & $<200\text{ns}$ \cite{Camenzind_spinHex} \\
    \hline
    Bias & $> 10$ & $> 10$ & $> 2500$ & $> 50$ \\
    \hline
    \end{tabular}%
    }
\end{table}

As shown above, hole-spin architectures like Ge/SiGe(p) generally exhibit a more pronounced noise bias compared to electron-spin variants. This is intimately linked to the strong spin-orbit coupling inherent to hole spins \cite{Bosco2023_shuttlingSOC, Ademi2025_shuttling}.

\subsection{RESULTS}
As summarized in Table \ref{tab:combined_thresholds}, the simulated threshold values, $p_{th} \times 10^3$, vary significantly depending on the code type, shuttling configuration, and the dominant noise bias. For the CSS code under purely symmetric noise ($\eta=0.5$, $\eta_s=0.5$), the threshold is approximately $2.2 \times 10^{-3}$ for data shuttling and $2.65 \times 10^{-3}$ for check shuttling. As we introduce higher X-bias ($\eta=100$, $\eta_s=100$), the CSS check shuttling configuration exhibits a highly asymmetric memory performance: the Z memory threshold exceeds $5 \times 10^{-3}$, while the X memory remains limited to roughly $2.45 \times 10^{-3}$. Since the functional threshold is dictated by the minimum of the two bases, the overall CSS threshold cannot surpass $\sim 2.45 \times 10^{-3}$ under these strongly biased conditions.
In contrast, the XZZX surface code demonstrates a more balanced threshold enhancement under biased noise, particularly when the shuttling noise is Z-biased ($\eta_Z$). For the XZZX data shuttling configuration, the threshold increases from $2.2 \times 10^{-3}$ at symmetric noise to $3.5 \times 10^{-3}$ at maximal bias ($\eta=100$, $\eta_s=100$). The highest overall threshold is achieved using the XZZX code in a check shuttling configuration under Z-bias, reaching up to $4.0 \times 10^{-3}$ when both gate and shuttling noise are highly biased ($\eta=100$, $\eta_s=100$). This configuration effectively exploits the bias-tailored structure of the XZZX code to deliver optimal fault-tolerant performance in the railway architecture. We have simulated the extraction process under realistic circuit level noise. The thresholds of V and H memory for XZZX RSC and Z and X memory for CSS RSC has been calculated. Since we would like to operate in either basis, we take the minimum of their values to represent the performance of the code under bias.
\begin{equation}p_{th}=min\{p_{th}^H,p_{th}^V\}\end{equation}

Following the threshold estimations presented above, we now turn our attention to the physical qubit overhead required to achieve practical fault-tolerant computation. These footprint evaluations are in the following section below.

\begin{figure}
    \centering
    \begin{tikzpicture}
        \node[anchor=south west,inner sep=0] (image) at (0,0) {\includegraphics[width=\linewidth]{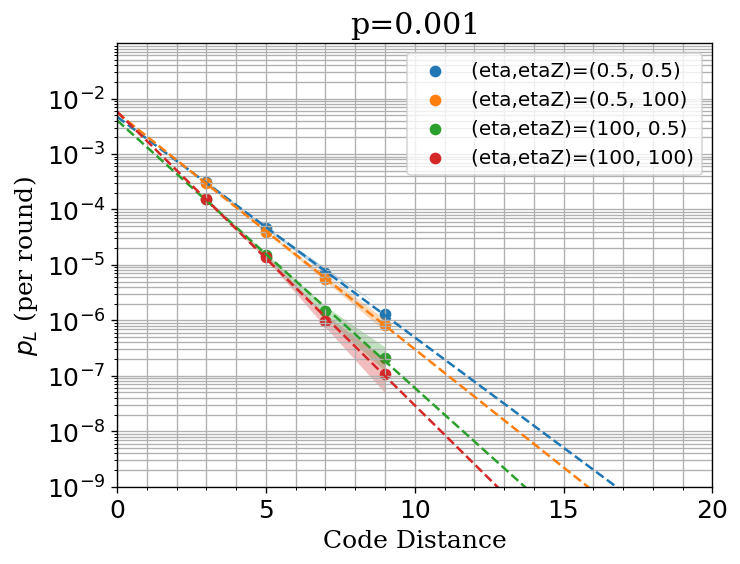}};
        \begin{scope}[x={(image.south east)},y={(image.north west)}]
            \node[font=\large, color=black] at (0.05, 0.95) {a)};
        \end{scope}
    \end{tikzpicture}

\vspace{0.5cm}
    \begin{tikzpicture}
        \node[anchor=south west,inner sep=0] (image) at (0,0) {\includegraphics[width=\linewidth]{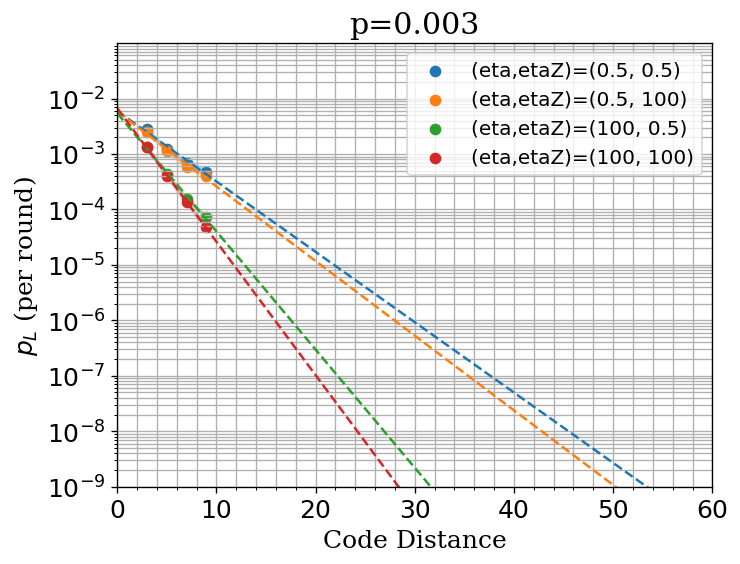}};
        \begin{scope}[x={(image.south east)},y={(image.north west)}]
            \node[font=\large, color=black] at (0.05, 0.95) {b)};
        \end{scope}
    \end{tikzpicture}
    \vspace{-1.5em}

\caption{The logical error rates per round versus the rotated XZZX surface-code distances with the extraction circuit and shuttling pattern. The logical error rates in these figures are for a horizontal (H) memory experiment. Both memories behave similarly deep below threshold, with the H memory showing a slightly worse logical error probability, $p_L$, and thus it is the limiting case. (a) Physical error rate $p = 0.001$. (b) Physical error rate $p = 0.003$.}
\label{fig:p1megaquop}
\label{fig:p3megaquop}
\end{figure}

\section{VI. Footprints}
We now discuss the required qubit footprints to reach certain regimes of error-free quantum operations as a function of the bias. We have numerically computed the logical error rates per round, $p_L$, as a function of the code distance. We examine two experimentally relevant physical error rates: $p = 0.003$ and $p = 0.001$. As observed in Fig. \ref{fig:p1megaquop}, the required distances to reach certain logical error probabilities are significantly reduced when a biased system is considered. Furthermore, the reduction in logical error rate provided by the bias exhibits a saturation effect for $\eta \geq 100$. We highlight error bars showing values for which the conditional probabilities $P(p_L|k)$ are within a factor of $1000$ of the maximum-likelihood estimation, $p_L = k/n$, assuming a binomial distribution to account for statistical uncertainties \cite{PhysRevResearch.7.033074}.
Table \ref{tab:footprints_combined} quantifies the qubit footprints required to reach the Megaquop and Gigaquop regimes at a physical error rate of $p = 0.003$. The reduction in qubit footprint ranges from around $64\%$ for a purely idling bias of $\eta=100$ to around $72\%$ when the shuttling bias is also maximized ($\eta=100, \eta_Z=100$), compared to the baseline symmetric $(\eta=0.5, \eta_Z=0.5)$ noise model. Notably, the idling and CZ bias $\eta$ plays a significantly more dominant role than shuttling bias $\eta_Z$ in reducing the required footprint, as observed by the stark difference in reduction between $(100, 0.5)$ and $(0.5, 100)$.
Similarly, Table \ref{tab:footprints_combined} quantifies the footprint requirements for the $p = 0.001$ physical error rate case. Inspecting the table, we observe that for this lower physical error rate, the decrease in qubit footprint ranges from $42\%$ up to around $51\%$.
The main conclusion that can be drawn from these results is that noise bias toward dephasing errors can be powerfully leveraged by XZZX surface codes tailored to the $2N$ array architecture, resulting in massive reductions in the required number of physical qubits. These significant footprint reductions come ``for free" for qubit platforms where bias-preserving CZ gates and shuttling operations are natively enabled. We note that the shuttling protocol is not specific to the rotated surface codes and other stabilizer codes in general can be error corrected in this shuttling protocol if a suitable mapping can be obtained for them.

% \bl{Need comparison with CSS? Will need simulations to be run again for 3,5,7,9 distances. Also only for eta/etaX not eta/etaZ. So better not?}
% XZZX Megaquop footprint gives a distance of $7$ for check qubit shuttled under Z bias noise.
% For other configurations, the distance $8$ is suitable.
% This gives and overall physical qubit count of $97$ and $127$ respectively, for achieving $10^{-6}$ logical error rate.
% \bl{}

\section{VII. Conclusion}
We introduced a mapping of the rotated surface codes (CSS and XZZX) to $2\times N$ array. We also proposed an optimized shuttling protocol, which minimizes idling and shuttling instances. These were designed to combat hook errors and satisfy the commutation constraints of the stabilizers of the code. The performance of the codes in the $2N$ array architecture under circuit level noise has been studied. The adopted noise model with varying degrees of bias, is realistic for spin qubit architectures \cite{Veldhorst2017_CMOSarchitecture, Li2018_crossbar, Brousse2022_dense2D, li2025trilinear, Escofet2025Quantum, Chadwick2025_SNAQ}.
As seen in the table \ref{tab:combined_thresholds}, XZZX code is able to leverage the biases, especially the idling bias, to improve the performance over regular CSS rotated surface code. On top of that, shuttling the check qubits rather than the data qubits provide a significant improvement in performance. We hypothesize that this is due to the fact that idling the data qubits (and shuttling the check qubits) preserve the noise bias (Z), which is advantageous for XZZX code. We also note a slight improvement in threshold when the shuttling bias is Z instead of X. This could be because the Z errors do not get transmitted down to the data qubits, and could be easily corrected as measurement errors in the subsequent rounds. Across a wide range of asymmetry between shuttling-induced errors and the idling/CZ background, our simulations show that non-CSS structure tailored to biased noie by Clifford deformations can systematically enhance robustness when the dominant error mechanisms are aligned with the code’s native error-suppression pattern. Moreover, we find that shuttling the check qubits yields consistently higher thresholds than moving data qubits, highlighting a concrete shuttling strategy to improve code performance in constrained railway geometries.

Under depolarizing noise (no bias), the configuration of shuttling checks still give better performance for both XZZX and CSS rotated surface codes. The CSS RSC performs better than the XZZX in this case, whereas under biased noise, XZZX gives the better performance. We hypothesize that with idling bias of $0.5$, the 2 qubit gate noise is not depolarizing, and is giving a slight advantage to the CSS code. Where as, with bias $>> 1$, the logical operators (undetectable errors) of XZZX code becomes very unlikely. This validates our intuition to use XZZX to leverage the bias when it is present in the noise.
It is important to emphasize that from Fig. \ref{fig:p1megaquop}, at a physical error rate of $p=0.001$ and a bias of $(\eta=100,\eta_Z=100)$, one could reach the Megaquop footprint with a code distance of just $d=7$. This is a crucial result, as it indicates that the necessary railway geometry processor will have a very moderate size. Specifically, a distance $7$ rotated surface code requires exactly $97$ physical qubits ($49$ data and $48$ check qubits) in total to reach the Megaquop regime, making such a device highly viable for near-term implementations of quantum fault-tolerance.
Furthermore, the flexibility of this bilinear architecture extends beyond single logical qubits. The integration of a third rail opens up a direct pathway for executing transversal gates and interacting with adjacent logical qubit systems.

Crucially, while concurrent works have explored early fault-tolerance on shuttling-based 2xN arrays \cite{Siegel2024Towards} and optimal 1D surface-code topologies \cite{Escofet2025Quantum, Chadwick2025_SNAQ}, our results uniquely highlight the interplay between restrictive linear geometries and the underlying physical noise bias. Specifically, we demonstrate that tailoring the topological code choice (XZZX vs. CSS) and the shuttling strategy to the predominant error mechanisms of semiconductor qubits becomes essential for maximizing the fault-tolerant footprint reduction in these highly constrained platforms.

\FloatBarrier
\section*{Acknowledgments}
 This work was supported by the Ministry for Digital Transformation and of Civil Service of the Spanish Government through the QUANTUM ENIA project call---QUANTUM SPAIN project, and by the European Union through the Recovery, Transformation and Resilience Plan---NextGenerationEU within the framework of the Digital Spain 2026 Agenda. We also acknowledge support from the Basque Quantum (BasQ) strategy.

\bibliography{references}

\end{document}

%% file: figures/2N.tex
\begin{tikzpicture}[scale=.7]
\begin{scope}[yshift=4cm]
  % \node at (5,.7) {\uline{Format 1}};
  % Horizontal lines
  % \draw (-.1,0) -- (10.1,0);
  \draw (0,-0.8) -- (10,-0.8);
  \draw (0,-1.6) -- (10,-1.6);

  % Dots on the first line (black!75 and white)
  \foreach \i in {0, 0.5, 1, 1.5, 2, 2.5, 3, 3.5, 4, 4.5, 5, 5.5, 6, 6.5, 7, 7.5, 8, 8.5, 9, 9.5, 10} {
    \draw[thin, gray!30] (\i,-.8)--(\i,-1.6);
    \pgfmathsetseed{int(100*\i)}
    \pgfmathsetmacro{\randomColor}{rnd}

    % \ifdim\randomColor pt < 0.5pt
    %   \node[circle,draw,fill=black!75, inner sep=1pt, minimum size=7pt] at (\i,0){};
    % \else
      % \node[circle,draw,fill=white, inner sep=1pt, minimum size=7pt] at (\i,0){};
    % \fi
  }

  % Dots on the second line (red, blue, and white)
  \foreach \i in {0, 0.5, 1, 1.5, 2, 2.5, 3, 3.5, 4, 4.5, 5, 5.5, 6, 6.5, 7, 7.5, 8, 8.5, 9, 9.5, 10} {
  
    \pgfmathsetseed{int(100*\i + 1000)}
    \pgfmathsetmacro{\randomColor}{rnd}

    \ifdim\randomColor pt < 0.33pt
      \node[circle,draw,fill=white, inner sep=1pt, minimum size=7pt] at (\i,-.8){};
      \node[circle,draw,fill opacity=0.5,fill=gray, inner sep=1pt, minimum size=7pt] at (\i,-.8){};
    \else\ifdim\randomColor pt < 0.66pt
      \node[circle,draw,fill=white, inner sep=1pt, minimum size=7pt] at (\i,-.8){};
      \node[circle,draw,fill opacity=0.8,fill=darkgray, inner sep=1pt, minimum size=7pt] at (\i,-.8){};
    \else
      \node[circle,draw,fill=white, inner sep=1pt, minimum size=7pt] at (\i,-.8){};
    \fi\fi
  }
[thick,fill opacity=0.18,fill=gray][thick,fill opacity=0.3,fill=darkgray]
  % Dots on the third line (black!75 and white) - Same as the first line
  \foreach \i in {0, 0.5, 1, 1.5, 2, 2.5, 3, 3.5, 4, 4.5, 5, 5.5, 6, 6.5, 7, 7.5, 8, 8.5, 9, 9.5, 10} {
    \pgfmathsetseed{int(100*\i + 2000)}
    \pgfmathsetmacro{\randomColor}{rnd}

    \ifdim\randomColor pt < 0.5pt
      \node[circle,draw,fill=black!75, inner sep=1pt, minimum size=7pt] at (\i,-1.6){};
    \else
      \node[circle,draw,fill=white, inner sep=1pt, minimum size=7pt] at (\i,-1.6){};
    \fi
  }
\end{scope}
\end{tikzpicture}

%% file: figures/XZZXd5.tex
\begin{tikzpicture}
\begin{scope}

\foreach \x in {0,1,2,3} {
\ifodd\numexpr\x\relax
        \draw[thick,fill opacity=0.3,fill=darkgray] (\x,0) --(\x+1,0)--(\x+.5,-.5)--cycle;
        \draw[thick,fill opacity=0.18,fill=gray] (4,\x) --(4,\x+1)--(4.5,\x+.5)--cycle; %pattern=checkerboard light gray]
\else
        \draw[thick,fill opacity=0.3,fill=darkgray] (\x,4) --(\x+1,4)--(\x+.5,4.5)--cycle;%pattern=dots]
        
        \draw[thick,fill opacity=0.18,fill=gray]  (0,\x) --(0,\x+1)--(-.5,\x+.5)--cycle;
        \fi
}

 \foreach \x in {0,1,2,3} {
    \foreach \y in {0,1,2,3} {
      \ifodd\numexpr\x+\y\relax
        % \fill[soothingblue!75!soothingred] (\x,\y) rectangle +(1,1);
        \draw[thick,fill opacity=0.18,fill=gray]  (\x,\y) rectangle +(1,1);
        % \fill[soothingred!75!soothingblue] (\x,\y+1) -- (\x+1,\y+1) -- (\x+.5, \y+1.5) -- cycle;
        % \fill[soothingred!75!soothingblue] (\x,\y) -- (\x+1,\y) -- (\x+.5, \y-.5) -- cycle;
       % \draw (\x,\y+1) -- (\x+1,\y+1) -- (\x+.5, \y+1.5) -- cycle;
       % \draw[gray] (\x,\y) -- (\x+1,\y) -- (\x+.5, \y-.5) -- cycle;
       
      \else
        \draw[thick,fill opacity=0.3,fill=darkgray] (\x,\y) rectangle +(1,1);
        % \fill[soothingred!75!soothingblue] (\x,\y) rectangle +(1,1);
        % \fill[soothingblue!75!soothingred] (\x,\y) -- (\x,\y+1) -- (\x-.5,\y+.5) -- cycle;
        % \fill[soothingblue!75!soothingred] (\x+1,\y) -- (\x+1,\y+1) -- (\x+1.5,\y+.5) -- cycle;
       % \draw (\x,\y) -- (\x,\y+1) -- (\x-.5,\y+.5) -- cycle;
       % \draw (\x+1,\y) -- (\x+1,\y+1) -- (\x+1.5,\y+.5) -- cycle;
      \fi
    }
    
  }
   \foreach \x in {0,1,2,3} {
    \foreach \y in {0,1,2,3} {
      \ifodd\numexpr\x+\y\relax
      
       \fi
    }
  }

  \node at (.17,3.83) (a){\tiny X};
  \node at (.83,3.83) (b){\tiny Z};
  \node at (.17,3.17) (c){\tiny Z};
  \node at (.83,3.17) (d){\tiny X};
  \draw[thin] (a)--(b)--(d)--(c)--(a);
    
  \node at (.17,2.83) (a){\tiny X};
  \node at (.83,2.83) (b){\tiny Z};
  \node at (.17,2.17) (c){\tiny Z};
  \node at (.83,2.17) (d){\tiny X};
  \draw[thin] (a)--(b)--(d)--(c)--(a);

  \node at (-.1,2.75) (a) {\tiny Z};
  \node at (-.1,2.25) (b) {\tiny X};
  \draw[thin] (a)--(b);
  
   \node at (4+.1,1+2.75) (a) {\tiny X};
  \node at (4+.1,1+2.25) (b) {\tiny Z};
  \draw[thin] (a)--(b);
  
   \node at (.75,4+.1) (a) {\tiny X};
  \node at (.25,4+.1) (b) {\tiny Z};
  \draw[thin] (a)--(b);
  
   \node at (1.75,-.105) (a) {\tiny Z};
  \node at (1.25,-.105) (b) {\tiny X};
  \draw[thin] (a)--(b);
  
  \end{scope}
\end{tikzpicture}

%% file: figures/XZZXSyndromeExtractionCircuit.tex
\begin{tikzpicture}[scale=.8]

    \clip (-1,-1.8) rectangle (10,3.5);
    % Data qubits
    \node[inner sep=5pt] at (0.7,0) (q0) {$i$};
    \node[inner sep=5pt] at (0.7,1) (q1) {$j$};
    \node[inner sep=5pt] at (0.7,2) (q2) {$k$};
    \node[inner sep=5pt] at (0.7,3) (q3) {$l$};
    \node[inner sep=5pt] at (0.7,-1) (ancilla) {$\ket{0}$};
    % Horizontal lines for qubits
    \draw[thick] (q0.east) -- ++(8,0);
    \draw[thick] (ancilla.east) -- ++(7.3,0);
    \draw[thick] (q1.east) -- ++(8,0);
    \draw[thick] (q2.east) -- ++(8,0);
    \draw[thick] (q3.east) -- ++(8,0);

    % CNOT Gates: control on q0 and target on ancilla
    \draw[fill=black] (3,-1) circle [radius=0.1]; % CNOT control on q0
    \draw[thick] (3, 0) circle [radius=0.2]; % CNOT target on ancilla
    \draw[thick] (3, -1) -- (3, 0); % Vertical connection line

    % CNOT Gates: control on q1 and target on ancilla
    \draw[fill=black] (4,-1) circle [radius=0.1]; % CNOT control on q0
    % \draw[thick] (4, 1) circle [radius=0.2]; % CNOT target on ancilla
    \draw[fill=black] (4,1) circle [radius=0.1]; % CNOT control on q0
    \draw[thick] (4, -1) -- (4, 1); % Vertical connection line

    % CNOT Gates: control on q2 and target on ancilla
    \draw[fill=black] (5,-1) circle [radius=0.1]; % CNOT control on q0
    \draw[fill=black] (5,2) circle [radius=0.1]; % CNOT control on q0
    % \draw[thick] (5, 2) circle [radius=0.2]; % CNOT target on ancilla
    \draw[thick] (5, -1) -- (5, 2); % Vertical connection line
    
    % CNOT Gates: control on q3 and target on ancilla
    \draw[fill=black] (6,-1) circle [radius=0.1]; % CNOT control on q0
    \draw[thick] (6, 3) circle [radius=0.2]; % CNOT target on ancilla
    \draw[thick] (6, -1) -- (6, 3); % Vertical connection line
    
  \draw[thick, fill=white] (6.5+.5,-1.3) rectangle (7.0+.5,-0.7);
    \node at (6.75+.5,-1) {$H$};

    \draw[thick, fill=white] (6.5-5,-1.3) rectangle (7.0-5,-0.7);
    \node at (6.75-5,-1) {$H$};

    \node at (0.7,-1.6) {``check qubit"};
    
\begin{scope}[xshift=1.5cm]
    % Measurement symbol on ancilla
    \draw[thick,fill=white] (6.75,-1.3) rectangle (7.25,-0.7);
%    \node at (7,-1) {$M$}; % Label for measurement
	\draw (7.18,-.98) arc (30:150:.2);
	\draw[->] (7,-1.1)--(7.15,-.8);
%    \node at (8,0.5) {$\ket{\psi}$};
%    \draw [decorate,decoration={brace,amplitude=5pt,mirror,raise=4ex}]
  (6.5,0) -- (6.5,1) node[midway,yshift=-3em]{};
  \end{scope}
  
    \node[text=darkerred] at (2.5,-1) (X0) {};
    \node[text=darkerred] at (9.1,0) (X1) {};
    \node[text=darkerred] at (9.1,1) (X2) {};
    \node[text=darkerred] at (9.1,2) (X3) {};
    \node[text=darkerred] at (9.1,3) (X4) {};
%    \begin{scope}[yshift=-4.2cm]
%    \node[text=darkerred] at (0,1) (a) {};
%    \node[text=darkerred] at (0,0) (b) {};
%    \node[text=darkerred] at (1,1) (c) {};
%    \node[text=darkerred] at (1,0) (d) {};
%    \draw[white] (a)--(b)--(d)--(c)--(a);
%    \end{scope}
    
    \node[text=darkerred, fill=white, inner sep=0pt] at (4.5,-1) (X0) {X};
%    \node[text=darkerred] at (9.1,0) (X1) {X};
%    \node[text=darkerred] at (9.1,1) (X2) {X};
    \node[text=darkerred, fill=white, inner sep=5pt] at (9.1,2) (X3) {Z};
    \node[text=darkerred, fill=white, inner sep=5pt] at (9.1,3) (X4) {X};
    \end{tikzpicture}
    

%% file: figures/HookErrord5.tex
\begin{tikzpicture}
    
 \foreach \x in {0,1,2,3} {
    \foreach \y in {0,1,2,3} {
      \ifodd\numexpr\x+\y\relax
        \fill[soothingblue] (\x,\y) rectangle +(1,1);
        \fill[soothingred] (\x,\y+1) -- (\x+1,\y+1) -- (\x+.5, \y+1.5) -- cycle;
        \fill[soothingred] (\x,\y) -- (\x+1,\y) -- (\x+.5, \y-.5) -- cycle;
       % \draw (\x,\y+1) -- (\x+1,\y+1) -- (\x+.5, \y+1.5) -- cycle;
       % \draw (\x,\y) -- (\x+1,\y) -- (\x+.5, \y-.5) -- cycle;
       
      \else
        \fill[soothingred] (\x,\y) rectangle +(1,1);
        \fill[soothingblue] (\x,\y) -- (\x,\y+1) -- (\x-.5,\y+.5) -- cycle;
        \fill[soothingblue] (\x+1,\y) -- (\x+1,\y+1) -- (\x+1.5,\y+.5) -- cycle;
       % \draw (\x,\y) -- (\x,\y+1) -- (\x-.5,\y+.5) -- cycle;
       % \draw (\x+1,\y) -- (\x+1,\y+1) -- (\x+1.5,\y+.5) -- cycle;
      \fi
    }
  }
   \foreach \x in {0,1,2,3} {
    \foreach \y in {0,1,2,3} {
      \ifodd\numexpr\x+\y\relax
      
       \fi
    }
  }
  % Draw grid lines (optional, but makes it clearer)
 % \draw[gray!50] (0,0) grid (4,4);  
  \foreach \x in {0,1,2,3,4} {
        \foreach \y in {0,1,2,3,4} {
            % Node at each grid point
            \fill (\x, \y) circle (3pt);
            
            % Horizontal lines (except the last column)
            \ifnum\x<4
                \draw (\x, \y) -- ++(1,0);
            \fi
            
            % Vertical lines (except the last row)
            \ifnum\y<4
                \draw (\x, \y) -- ++(0,1);
            \fi
        }
    }
    \draw (0,4)--(0.5,4.5)--(1,4);
    \draw (2,4)--(2.5,4.5)--(3,4);
    \draw (1,0)--(1.5,-0.5)--(2,0);
    \draw (3,0)--(3.5,-0.5)--(4,0);
    \draw (0,0)--(-0.5,0.5)--(0,1);
    \draw (0,2)--(-0.5,2.5)--(0,3);
    \draw (4,3)--(4.5,3.5)--(4,4);
    \draw (4,1)--(4.5,1.5)--(4,2);

    \begin{scope}
    \draw[->, thick] (1.2,.8)--(1.8,.8)--(1.2,.2)--(1.8,.2);
    
    \node[circle, inner sep= .1em, fill=yellow] at (1,0) (z1) {X};
    \node[circle, inner sep= .1em, fill=yellow] at (2,0) (z2) {X};
    \draw[line width=1mm, yellow] (z1)--(z2);
  \end{scope}
  
    \begin{scope}[xshift=2cm]
    \draw[->, thick] (1.2,.8)--(1.8,.8)--(1.2,.2)--(1.8,.2);
    
    \node[circle, inner sep= .1em, fill=yellow] at (1,0) (z3) {X};
    \node[circle, inner sep= .1em, fill=yellow] at (2,0) (z4) {X};
    \draw[line width=1mm, yellow] (z3)--(z4);
  \end{scope}
  
    \begin{scope}
%    \draw[->, thick] (1.2,.8)--(1.8,.8)--(1.2,.2)--(1.8,.2);
    
    \node[circle, inner sep= .1em, fill=yellow,text=red] at (1.5,.5) {X};
    \node[circle, inner sep= .1em, fill=yellow, text=red] at (3.5,.5) {X};
    \node[circle, inner sep= .1em, fill=yellow,text=red] at (0,0) (z0) {X};
    \draw[line width=1mm, yellow] (z1)--(z0);
    \draw[line width=1mm, yellow] (z2)--(z3);
  \end{scope}

    \node at (1.17,3.83) (a){\tiny Z};
  \node at (1.83,3.83) (b){\tiny Z};
  \node at (1.17,3.17) (c){\tiny Z};
  \node at (1.83,3.17) (d){\tiny Z};
  \draw[thin] (a)--(b)--(d)--(c)--(a);
    
  \node at (2.17,3.83) (a){\tiny X};
  \node at (2.83,3.83) (b){\tiny X};
  \node at (2.17,3.17) (c){\tiny X};
  \node at (2.83,3.17) (d){\tiny X};
  \draw[thin] (a)--(b)--(d)--(c)--(a);
\end{tikzpicture}

%% file: figures/XZZXplaquettes.tex
\begin{tikzpicture}[scale=0.5]

\begin{scope}[scale=0.6,xshift=-7cm,yshift=1cm]
\fill[thick,fill opacity=0.18,fill=gray] (0,0)--(1,0)--(1,1)--(0,1)--(0,0);
\end{scope}

\begin{scope}[scale=0.6,xshift=-7cm,yshift=3cm]
\fill[thick,fill opacity=0.18,fill=gray] (0,0)--(1,0)--(1,1)--(0,1)--(0,0);
\end{scope}

\begin{scope}[scale=0.6,xshift=-7cm,yshift=5cm]
\fill[thick,fill opacity=0.18,fill=gray] (0,0)--(1,0)--(1,1)--(0,1)--(0,0);
\end{scope}

\begin{scope}[scale=0.6,xshift=-6cm,yshift=0cm]
\fill[thick,fill opacity=0.18,fill=gray] (0,0)--(1,0)--(1,1)--(0,1)--(0,0);
\end{scope}

\begin{scope}[scale=0.6,xshift=-6cm,yshift=2cm]
\fill[thick,fill opacity=0.18,fill=gray] (0,0)--(1,0)--(1,1)--(0,1)--(0,0);
\end{scope}

\begin{scope}[scale=0.6,xshift=-6cm,yshift=4cm]
\fill[thick,fill opacity=0.18,fill=gray] (0,0)--(1,0)--(1,1)--(0,1)--(0,0);
\end{scope}

\begin{scope}[scale=0.6,xshift=-5cm,yshift=1cm]
\fill[thick,fill opacity=0.18,fill=gray] (0,0)--(1,0)--(1,1)--(0,1)--(0,0);
\end{scope}

\begin{scope}[scale=0.6,xshift=-5cm,yshift=3cm]
\fill[thick,fill opacity=0.18,fill=gray] (0,0)--(1,0)--(1,1)--(0,1)--(0,0);
\end{scope}

\begin{scope}[scale=0.6,xshift=-5cm,yshift=5cm]
\fill[thick,fill opacity=0.18,fill=gray] (0,0)--(1,0)--(1,1)--(0,1)--(0,0);
\end{scope}

\begin{scope}[scale=0.6,xshift=-4cm,yshift=0cm]
\fill[thick,fill opacity=0.18,fill=gray] (0,0)--(1,0)--(1,1)--(0,1)--(0,0);
\end{scope}

\begin{scope}[scale=0.6,xshift=-4cm,yshift=2cm]
\fill[thick,fill opacity=0.18,fill=gray] (0,0)--(1,0)--(1,1)--(0,1)--(0,0);
\end{scope}

\begin{scope}[scale=0.6,xshift=-4cm,yshift=4cm]
\fill[thick,fill opacity=0.18,fill=gray] (0,0)--(1,0)--(1,1)--(0,1)--(0,0);
\end{scope}

\begin{scope}[scale=.6,xshift=-7cm,yshift=11cm]
\fill[thick,fill opacity=0.3,fill=darkgray] (0,0)--(1,0)--(1,1)--(0,1)--(0,0);
\end{scope}

\begin{scope}[scale=.6,xshift=-5cm,yshift=11cm]
\fill[thick,fill opacity=0.3,fill=darkgray] (0,0)--(1,0)--(1,1)--(0,1)--(0,0);
\end{scope}

\begin{scope}[scale=.6,xshift=-3cm,yshift=11cm]
\fill[thick,fill opacity=0.3,fill=darkgray] (0,0)--(1,0)--(1,1)--(0,1)--(0,0);
\end{scope}

\begin{scope}[scale=.6,xshift=-8cm,yshift=10cm]
\fill[thick,fill opacity=0.3,fill=darkgray] (0,0)--(1,0)--(1,1)--(0,1)--(0,0);
\end{scope}

\begin{scope}[scale=.6,xshift=-6cm,yshift=10cm]
\fill[thick,fill opacity=0.3,fill=darkgray] (0,0)--(1,0)--(1,1)--(0,1)--(0,0);
\end{scope}

\begin{scope}[scale=.6,xshift=-4cm,yshift=10cm]
\fill[thick,fill opacity=0.3,fill=darkgray] (0,0)--(1,0)--(1,1)--(0,1)--(0,0);
\end{scope}

\begin{scope}[scale=.6,xshift=-7cm,yshift=9cm]
\fill[thick,fill opacity=0.3,fill=darkgray] (0,0)--(1,0)--(1,1)--(0,1)--(0,0);
\end{scope}

\begin{scope}[scale=.6,xshift=-5cm,yshift=9cm]
\fill[thick,fill opacity=0.3,fill=darkgray] (0,0)--(1,0)--(1,1)--(0,1)--(0,0);
\end{scope}

\begin{scope}[scale=.6,xshift=-3cm,yshift=9cm]
\fill[thick,fill opacity=0.3,fill=darkgray] (0,0)--(1,0)--(1,1)--(0,1)--(0,0);
\end{scope}

\begin{scope}[scale=.6,xshift=-8cm,yshift=8cm]
\fill[thick,fill opacity=0.3,fill=darkgray] (0,0)--(1,0)--(1,1)--(0,1)--(0,0);
\end{scope}

\begin{scope}[scale=.6,xshift=-6cm,yshift=8cm]
\fill[thick,fill opacity=0.3,fill=darkgray] (0,0)--(1,0)--(1,1)--(0,1)--(0,0);
\end{scope}

\begin{scope}[scale=.6,xshift=-4cm,yshift=8cm]
\fill[thick,fill opacity=0.3,fill=darkgray] (0,0)--(1,0)--(1,1)--(0,1)--(0,0);
\end{scope}

%---------------------------------------
%---------------------------------------
\begin{scope}[xshift=0cm,yshift=0cm]
\fill[yellow!20] (-.3,-.3)--(3.3,-.3)--(3.3,8.3)--(-.3,8.3)--(-.3,-.3);
  \coordinate (a) at (0,1);
  \coordinate (b) at (1,1);
  \coordinate (c) at (0,0);
  \coordinate (d) at (1,0);
\draw[->, thick] (c)--(d)--(b)--(a);
\end{scope}
  
\begin{scope}[xshift=2cm,yshift=0cm]
  \coordinate (a) at (0,1);
  \coordinate (b) at (1,1);
  \coordinate (c) at (0,0);
  \coordinate (d) at (1,0);
\draw[->, thick] (a)--(b)--(d)--(c);
\end{scope}

\begin{scope}[xshift=10cm,yshift=0cm]
\fill[green!20] (-.3,-.3)--(3.3,-.3)--(3.3,8.3)--(-.3,8.3)--(-.3,-.3);
  \coordinate (a) at (0,1);
  \coordinate (b) at (1,1);
  \coordinate (c) at (0,0);
  \coordinate (d) at (1,0);
\draw[->, thick] (c)--(d)--(a)--(b);
\end{scope}
\begin{scope}[xshift=12cm,yshift=0cm]
  \coordinate (a) at (0,1);
  \coordinate (b) at (1,1);
  \coordinate (c) at (0,0);
  \coordinate (d) at (1,0);
\draw[->, thick] (d)--(c)--(b)--(a);
\end{scope}

\draw[dashed] (-.5,4)--(13.5,4);

\begin{scope}[xshift=5cm,yshift=0cm]
\fill[orange!20] (-.3,-.3)--(3.3,-.3)--(3.3,8.3)--(-.3,8.3)--(-.3,-.3);

\draw[->, very thick] (1.48,3.9)--(1.48,4.5);

\draw[->, very thick] (1.62,4.2)--(1.62,3.6);

  \coordinate (a) at (0,1);
  \coordinate (b) at (1,1);
  \coordinate (c) at (0,0);
  \coordinate (d) at (1,0);
\draw[->, thick] (a)--(d)--(c)--(b);
\end{scope}
  
\begin{scope}[xshift=7cm,yshift=0cm]
  \coordinate (a) at (0,1);
  \coordinate (b) at (1,1);
  \coordinate (c) at (0,0);
  \coordinate (d) at (1,0);
\draw[->, thick] (d)--(a)--(c)--(b);
\end{scope}
 %---------------------------------------
\begin{scope}[xshift=0cm,yshift=2cm]
  \coordinate (a) at (0,1);
  \coordinate (b) at (1,1);
  \coordinate (c) at (0,0);
  \coordinate (d) at (1,0);
\draw[->, thick] (b)--(a)--(c)--(d);
\end{scope}
  
\begin{scope}[xshift=2cm,yshift=2cm]
  \coordinate (a) at (0,1);
  \coordinate (b) at (1,1);
  \coordinate (c) at (0,0);
  \coordinate (d) at (1,0);
\draw[->, thick] (d)--(c)--(a)--(b);
\end{scope}
  
\begin{scope}[xshift=5cm,yshift=2cm]
  \coordinate (a) at (0,1);
  \coordinate (b) at (1,1);
  \coordinate (c) at (0,0);
  \coordinate (d) at (1,0);
\draw[->, thick] (a)--(d)--(b)--(c);
\end{scope}
  
\begin{scope}[xshift=7cm,yshift=2cm]
  \coordinate (a) at (0,1);
  \coordinate (b) at (1,1);
  \coordinate (c) at (0,0);
  \coordinate (d) at (1,0);
\draw[->, thick] (d)--(a)--(b)--(c);
\end{scope}
  
\begin{scope}[xshift=10cm,yshift=2cm]
  \coordinate (a) at (0,1);
  \coordinate (b) at (1,1);
  \coordinate (c) at (0,0);
  \coordinate (d) at (1,0);
\draw[->, thick] (a)--(b)--(c)--(d);
\end{scope}
\begin{scope}[xshift=12cm,yshift=2cm]
  \coordinate (a) at (0,1);
  \coordinate (b) at (1,1);
  \coordinate (c) at (0,0);
  \coordinate (d) at (1,0);
\draw[->, thick] (b)--(a)--(d)--(c);
\end{scope} 

%------------------------------------------
%------------------------------------------

\begin{scope}[xshift=0cm,yshift=5cm]
  \coordinate (a) at (0,1);
  \coordinate (b) at (1,1);
  \coordinate (c) at (0,0);
  \coordinate (d) at (1,0);
\draw[->, thick] (c)--(a)--(b)--(d);
\end{scope}
  
\begin{scope}[xshift=2cm,yshift=5cm]
  \coordinate (a) at (0,1);
  \coordinate (b) at (1,1);
  \coordinate (c) at (0,0);
  \coordinate (d) at (1,0);
\draw[->, thick] (d)--(b)--(a)--(c);
\end{scope}
  
\begin{scope}[xshift=5cm,yshift=5cm]
  \coordinate (a) at (0,1);
  \coordinate (b) at (1,1);
  \coordinate (c) at (0,0);
  \coordinate (d) at (1,0);
\draw[->, thick] (b)--(c)--(a)--(d);
\end{scope}
  
\begin{scope}[xshift=7cm,yshift=5cm]
  \coordinate (a) at (0,1);
  \coordinate (b) at (1,1);
  \coordinate (c) at (0,0);
  \coordinate (d) at (1,0);
\draw[->, thick] (c)--(b)--(a)--(d);
\end{scope}
  
\begin{scope}[xshift=10cm,yshift=5cm]
  \coordinate (a) at (0,1);
  \coordinate (b) at (1,1);
  \coordinate (c) at (0,0);
  \coordinate (d) at (1,0);
\draw[->, thick] (c)--(a)--(d)--(b);
\end{scope}
\begin{scope}[xshift=12cm,yshift=5cm]
  \coordinate (a) at (0,1);
  \coordinate (b) at (1,1);
  \coordinate (c) at (0,0);
  \coordinate (d) at (1,0);
\draw[->, thick] (d)--(b)--(c)--(a);
\end{scope} 
%-------------------------------------------
\begin{scope}[xshift=0cm,yshift=7cm]
  \coordinate (a) at (0,1);
  \coordinate (b) at (1,1);
  \coordinate (c) at (0,0);
  \coordinate (d) at (1,0);
\draw[->, thick] (a)--(c)--(d)--(b);
\end{scope}
  
\begin{scope}[xshift=2cm,yshift=7cm]
  \coordinate (a) at (0,1);
  \coordinate (b) at (1,1);
  \coordinate (c) at (0,0);
  \coordinate (d) at (1,0);
\draw[->, thick] (b)--(d)--(c)--(a);
\end{scope}
  
\begin{scope}[xshift=5cm,yshift=7cm]
  \coordinate (a) at (0,1);
  \coordinate (b) at (1,1);
  \coordinate (c) at (0,0);
  \coordinate (d) at (1,0);
\draw[->, thick] (b)--(c)--(d)--(a);
\end{scope}
  
\begin{scope}[xshift=7cm,yshift=7cm]
  \coordinate (a) at (0,1);
  \coordinate (b) at (1,1);
  \coordinate (c) at (0,0);
  \coordinate (d) at (1,0);
\draw[->, thick] (c)--(b)--(d)--(a);
\end{scope}
  
\begin{scope}[xshift=10cm,yshift=7cm]
  \coordinate (a) at (0,1);
  \coordinate (b) at (1,1);
  \coordinate (c) at (0,0);
  \coordinate (d) at (1,0);
\draw[->, thick] (a)--(c)--(b)--(d);
\end{scope}
  
\begin{scope}[xshift=12cm,yshift=7cm]
  \coordinate (a) at (0,1);
  \coordinate (b) at (1,1);
  \coordinate (c) at (0,0);
  \coordinate (d) at (1,0);
\draw[->, thick] (b)--(d)--(a)--(c);
\end{scope}

% \draw[dashed] (-3,4)--(13.5,4);

%
\end{tikzpicture}

%% file: figures/XZZXcommutetrain.tex
\begin{tikzpicture}[scale=0.4]

% The plaquette of q_1 (a-c-b-d)
\begin{scope}[yshift=5cm]
    \fill[darkgray!30] (0,0)--(0,2)--(2,2)--(2,0)--(0,0);
    \draw[->, thick, gray] (.2,1.8)--(.2,.2)--(1.8,1.8)--(1.8,.2);
    \node at (1,1) {$q_1$}; % Adjust 0.1 for dot size
    \node[circle,fill=black!75, text=white, inner sep=1pt, minimum size=12pt] at (-.2,-.3) {k};
    \node[circle,fill=black!75, text=white, inner sep=1pt, minimum size=12pt] at (2.2,-.2) {l};
    \end{scope}
    \begin{scope}[yshift=7.6cm]
    \fill[gray!18] (0,0)--(0,2)--(2,2)--(2,0)--(0,0);
    \node at (1,1) {$q_2$}; % Adjust 0.1 for dot size
    \node[circle,fill=black!75, text=white, inner sep=1pt, minimum size=12pt] at (-.2,-.3) {i};
    \node[circle,fill=black!75, text=white, inner sep=1pt, minimum size=5pt] at (2.2,-.2) {j};
\end{scope}

% Rails + check qubits (red and blue) in the bottom rail
\begin{scope}
    \draw[->, thick] (16,3.5)--(15,3.5);
    \draw (0,0)--(20,0);
    \draw[dotted] (20,0)--(21,0);
    \draw[dotted] (-1,0)--(0,0);
    \draw (0,2)--(20,2);
    \draw[dotted] (20,2)--(21,2);
    \draw[dotted] (-1,2)--(0,2);

    \node[circle,draw,fill=darkgray!30, inner sep=1pt, minimum size=15pt] at (1.5,0){};
    \node[circle,draw,fill opacity=1,fill=gray!18, inner sep=1pt, minimum size=15pt] at (3,0){$q_2$};
    \node[circle,draw,fill=darkgray!30, inner sep=1pt, minimum size=15pt] at (4.5,0){};
    \node[circle,draw,fill=darkgray!30, inner sep=1pt, minimum size=15pt] at (6,0){$q_1$};
    \node[circle,draw,fill opacity=1,fill=gray!18, inner sep=1pt, minimum size=15pt] at (7.5,0){};
\end{scope}

% data qubits (black!75) in the top rail
\begin{scope}
    \node[circle,fill=black!75, text=white, inner sep=1pt, minimum size=15pt] at (12,2){i};
    \node[circle,fill=black!75, text=white, inner sep=1pt, minimum size=15pt] at (13.5,2){};
    \node[circle,fill=black!75, text=white, inner sep=1pt, minimum size=15pt] at (15,2){k};
    \node[circle,fill=black!75, text=white, inner sep=1pt, minimum size=15pt] at (16.5,2){j};
    \node[circle,fill=black!75, text=white, inner sep=1pt, minimum size=15pt] at (18,2){};
    \node[circle,fill=black!75, text=white, inner sep=1pt, minimum size=15pt] at (19.5,2){l};
% \node at (14,2) {a};
% \node at (15,2) {c};
% \node at (15.5,2) {b};
% \node at (17,2) {d};
\end{scope}

% Chart (a,b) -> (q_1, q_2)
	\begin{scope}[xshift=13cm, yshift=5.5cm]
	\fill[white] (0,0)--(6,0)--(6,4)--(0,4)--(0,0);
	\draw (2,4)--(2,0);
	\draw[dotted] (0,2)--(6,2);
	\node at (1,3) {i};
	\node at (1,1) {j};
	\node at (5,3) {$q_2$};
	\node at (3,3) {$q_1$};
	\node at (5,1) {$q_2$};
	\node at (3,1) {$q_1$};
	\end{scope}
\end{tikzpicture}

%% file: figures/plaquetted5.tex
\begin{tikzpicture}

\node[circle, fill=black!75, inner sep=2pt, text=white, minimum size=14pt] at (0,4)(a){a};
\node[circle, fill=black!75, inner sep=2pt, text=white, minimum size=14pt] at (1,4)(b){b};
\node[circle, fill=black!75, inner sep=2pt, text=white, minimum size=14pt] at (2,4)(c){c};
\node[circle, fill=black!75, inner sep=.5pt, text=white, minimum size=14pt] at (3,4){};
\node[circle, fill=black!75, inner sep=.5pt, text=white, minimum size=14pt] at (4,4){};
\node[circle, fill=black!75, inner sep=2pt, text=white, minimum size=14pt] at (0,3)(d){d};
\node[circle, fill=black!75, inner sep=2pt, text=white, minimum size=14pt] at (1,3)(e){e};
\node[circle, fill=black!75, inner sep=2pt, text=white, minimum size=14pt] at (2,3)(f){f};
\node[circle, fill=black!75, inner sep=.5pt, text=white, minimum size=14pt] at (3,3){};
\node[circle, fill=black!75, inner sep=.5pt, text=white, minimum size=14pt] at (4,3){};
\node[circle, fill=black!75, inner sep=2pt, text=white, minimum size=14pt] at (0,2)(g){g};
\node[circle, fill=black!75, inner sep=2pt, text=white, minimum size=14pt] at (1,2)(h){h};
\node[circle, fill=black!75, inner sep=.5pt, text=white, minimum size=14pt] at (2,2){};
\node[circle, fill=black!75, inner sep=.5pt, text=white, minimum size=14pt] at (3,2){};
\node[circle, fill=black!75, inner sep=.5pt, text=white, minimum size=14pt] at (4,2){};
\node[circle, fill=black!75, inner sep=2pt, text=white, minimum size=14pt] at (0,1)(i){i};
\node[circle, fill=black!75, inner sep=.5pt, text=white, minimum size=14pt] at (1,1){};
\node[circle, fill=black!75, inner sep=.5pt, text=white, minimum size=14pt] at (2,1){};
\node[circle, fill=black!75, inner sep=.5pt, text=white, minimum size=14pt] at (3,1){};
\node[circle, fill=black!75, inner sep=.5pt, text=white, minimum size=14pt] at (4,1){};
\node[circle, fill=black!75, inner sep=.5pt, text=white, minimum size=14pt] at (0,0){};
\node[circle, fill=black!75, inner sep=.5pt, text=white, minimum size=14pt] at (1,0){};
\node[circle, fill=black!75, inner sep=.5pt, text=white, minimum size=14pt] at (2,0){};
\node[circle, fill=black!75, inner sep=.5pt, text=white, minimum size=14pt] at (3,0){};
\node[circle, fill=black!75, inner sep=.5pt, text=white, minimum size=14pt] at (4,0){};
\begin{scope}[yshift=1cm]
\draw[->,thick] (.3,.7)--(.3,.3)--(.7,.7)--(.7,.3);
\end{scope}

\begin{scope}[yshift=1cm,xshift=2cm]
\draw[->,thick] (.3,.7)--(.3,.3)--(.7,.7)--(.7,.3);
\end{scope}

\begin{scope}[yshift=3cm,xshift=2cm]
\draw[->,thick] (.3,.7)--(.3,.3)--(.7,.7)--(.7,.3);
\end{scope}

\begin{scope}[yshift=2cm,xshift=1cm]
\draw[->,thick] (.3,.7)--(.3,.3)--(.7,.7)--(.7,.3);
\end{scope}

\begin{scope}[yshift=0cm,xshift=1cm]
\draw[->,thick] (.3,.7)--(.3,.3)--(.7,.7)--(.7,.3);
\end{scope}

\begin{scope}[yshift=2cm,xshift=3cm]
\draw[->,thick] (.3,.7)--(.3,.3)--(.7,.7)--(.7,.3);
\end{scope}

\begin{scope}[yshift=0cm,xshift=3cm]
\draw[->,thick] (.3,.7)--(.3,.3)--(.7,.7)--(.7,.3);
\end{scope}

\begin{scope}[yshift=3cm,xshift=0cm]
\draw[->,thick] (.3,.7)--(.3,.3)--(.7,.7)--(.7,.3);
\end{scope}
%----------------------------------------------------------
 \begin{scope}[xshift=0cm,yshift=0cm]
 \draw[->, thick] (.3,.7)--(.7,.7)--(.3,.3)--(.7,.3);
 \end{scope}
 
 \begin{scope}[xshift=0cm,yshift=2cm]
 \draw[->, thick] (.3,.7)--(.7,.7)--(.3,.3)--(.7,.3);
 \end{scope}
 
 \begin{scope}[xshift=1cm,yshift=1cm]
 \draw[->, thick] (.3,.7)--(.7,.7)--(.3,.3)--(.7,.3);
 \end{scope}
 
 \begin{scope}[xshift=1cm,yshift=3cm]
 \draw[->, thick] (.3,.7)--(.7,.7)--(.3,.3)--(.7,.3);
 \end{scope}
 
 \begin{scope}[xshift=2cm,yshift=0cm]
 \draw[->, thick] (.3,.7)--(.7,.7)--(.3,.3)--(.7,.3);
 \end{scope}
 
 \begin{scope}[xshift=2cm,yshift=2cm]
 \draw[->, thick] (.3,.7)--(.7,.7)--(.3,.3)--(.7,.3);
 \end{scope}
 
 \begin{scope}[xshift=3cm,yshift=1cm]
 \draw[->, thick] (.3,.7)--(.7,.7)--(.3,.3)--(.7,.3);
 \end{scope}
 
 \begin{scope}[xshift=3cm,yshift=3cm]
 \draw[->, thick] (.3,.7)--(.7,.7)--(.3,.3)--(.7,.3);
 \end{scope}
 \end{tikzpicture}
 

%% file: figures/Snake_order.tex
\begin{tikzpicture}[scale=0.8]
\begin{scope}%[xshift=6cm]
\draw [->,very thick] (0,4)--(0,3)--(1,4)--(2,4)--(1,3)--(0,2)--(0,1)--(1,2)--(2,3)--(3,4)--(4,4)--(3,3)--(2,2)--(1,1)--(0,0)--(1,0)--(2,1)--(3,2)--(4,3)--(4,2)--(3,1)--(2,0)--(3,0)--(4,1)--(4,0);
\node[fill=white] at (0,4.3){START};
\node[fill=white] at (4,-.3){STOP};
\end{scope}
\end{tikzpicture}

%% file: figures/Snake_indices.tex
\begin{tikzpicture}[scale=0.8]
%{1,3,4,10,11,2,5,9,12,15,6,8,13,18,20,7,14,17,21,24,15,16,22,23,25}}

\begin{scope}
\node[circle, fill=black!75,fill opacity=0, inner sep=2pt, text=white, minimum size=10pt] at (0,4.3){1};
\node[circle, fill=black!75,fill opacity=0, inner sep=2pt, text=white, minimum size=10pt] at (4,-.3){1};
\node[circle, fill=black!75, inner sep=2pt, text=white, minimum size=10pt] at (0,4){1};
\node[circle, fill=black!75, inner sep=2pt, text=white, minimum size=10pt] at (1,4){3};
\node[circle, fill=black!75, inner sep=2pt, text=white, minimum size=10pt] at (2,4){4};
\node[circle, fill=black!75, inner sep=.5pt, text=white, minimum size=10pt] at (3,4){10};
\node[circle, fill=black!75, inner sep=.5pt, text=white, minimum size=10pt] at (4,4){11};
\node[circle, fill=black!75, inner sep=2pt, text=white, minimum size=10pt] at (0,3){2};
\node[circle, fill=black!75, inner sep=2pt, text=white, minimum size=10pt] at (1,3){5};
\node[circle, fill=black!75, inner sep=2pt, text=white, minimum size=10pt] at (2,3){9};
\node[circle, fill=black!75, inner sep=.5pt, text=white, minimum size=10pt] at (3,3){12};
\node[circle, fill=black!75, inner sep=.5pt, text=white, minimum size=10pt] at (4,3){19};
\node[circle, fill=black!75, inner sep=2pt, text=white, minimum size=10pt] at (0,2){6};
\node[circle, fill=black!75, inner sep=2pt, text=white, minimum size=10pt] at (1,2){8};
\node[circle, fill=black!75, inner sep=.5pt, text=white, minimum size=10pt] at (2,2){13};
\node[circle, fill=black!75, inner sep=.5pt, text=white, minimum size=10pt] at (3,2){18};
\node[circle, fill=black!75, inner sep=.5pt, text=white, minimum size=10pt] at (4,2){20};
\node[circle, fill=black!75, inner sep=2pt, text=white, minimum size=10pt] at (0,1){7};
\node[circle, fill=black!75, inner sep=.5pt, text=white, minimum size=10pt] at (1,1){14};
\node[circle, fill=black!75, inner sep=.5pt, text=white, minimum size=10pt] at (2,1){17};
\node[circle, fill=black!75, inner sep=.5pt, text=white, minimum size=10pt] at (3,1){21};
\node[circle, fill=black!75, inner sep=.5pt, text=white, minimum size=10pt] at (4,1){24};
\node[circle, fill=black!75, inner sep=.5pt, text=white, minimum size=10pt] at (0,0){15};
\node[circle, fill=black!75, inner sep=.5pt, text=white, minimum size=10pt] at (1,0){16};
\node[circle, fill=black!75, inner sep=.5pt, text=white, minimum size=10pt] at (2,0){22};
\node[circle, fill=black!75, inner sep=.5pt, text=white, minimum size=10pt] at (3,0){23};
\node[circle, fill=black!75, inner sep=.5pt, text=white, minimum size=10pt] at (4,0){25};
\end{scope}
% \begin{scope}[xshift=6cm]
% \draw [->,very thick] (0,4)--(0,3)--(1,4)--(2,4)--(1,3)--(0,2)--(0,1)--(1,2)--(2,3)--(3,4)--(4,4)--(3,3)--(2,2)--(1,1)--(0,0)--(1,0)--(2,1)--(3,2)--(4,3)--(4,2)--(3,1)--(2,0)--(3,0)--(4,1)--(4,0);
% \node[fill=white] at (0,4.3){START};
% \node[fill=white] at (4,-.3){STOP};
% \end{scope}
\end{tikzpicture}

%% file: figures/XZZXallqubitorder.tex
\begin{tikzpicture}
%{1,3,4,10,11,2,5,9,12,15,6,8,13,18,20,7,14,17,21,24,15,16,22,23,25}}
\begin{scope}[xshift=0cm,yshift=3cm]
\fill[thick,fill opacity=0.3,fill=darkgray] (0,0)--(1,0)--(1,1)--(0,1)--(0,0);
\end{scope}

\begin{scope}[xshift=2cm,yshift=3cm]
\fill[thick,fill opacity=0.3,fill=darkgray] (0,0)--(1,0)--(1,1)--(0,1)--(0,0);
\end{scope}

\begin{scope}[xshift=4cm,yshift=3cm]
\fill[thick,fill opacity=0.3,fill=darkgray] (0,0)--(1,0)--(1,1)--(0,1)--(0,0);
\end{scope}

\begin{scope}[xshift=-1cm,yshift=2cm]
\fill[thick,fill opacity=0.3,fill=darkgray] (0,0)--(1,0)--(1,1)--(0,1)--(0,0);
\end{scope}

\begin{scope}[xshift=1cm,yshift=2cm]
\fill[thick,fill opacity=0.3,fill=darkgray] (0,0)--(1,0)--(1,1)--(0,1)--(0,0);
\end{scope}

\begin{scope}[xshift=3cm,yshift=2cm]
\fill[thick,fill opacity=0.3,fill=darkgray] (0,0)--(1,0)--(1,1)--(0,1)--(0,0);
\end{scope}

\begin{scope}[xshift=0cm,yshift=1cm]
\fill[thick,fill opacity=0.3,fill=darkgray] (0,0)--(1,0)--(1,1)--(0,1)--(0,0);
\end{scope}

\begin{scope}[xshift=2cm,yshift=1cm]
\fill[thick,fill opacity=0.3,fill=darkgray] (0,0)--(1,0)--(1,1)--(0,1)--(0,0);
\end{scope}

\begin{scope}[xshift=4cm,yshift=1cm]
\fill[thick,fill opacity=0.3,fill=darkgray] (0,0)--(1,0)--(1,1)--(0,1)--(0,0);
\end{scope}

\begin{scope}[xshift=-1cm,yshift=0cm]
\fill[thick,fill opacity=0.3,fill=darkgray] (0,0)--(1,0)--(1,1)--(0,1)--(0,0);
\end{scope}

\begin{scope}[xshift=1cm,yshift=0cm]
\fill[thick,fill opacity=0.3,fill=darkgray] (0,0)--(1,0)--(1,1)--(0,1)--(0,0);
\end{scope}

\begin{scope}[xshift=3cm,yshift=0cm]
\fill[thick,fill opacity=0.3,fill=darkgray] (0,0)--(1,0)--(1,1)--(0,1)--(0,0);
\end{scope}

% \begin{scope}%[xshift=6cm]
% \draw [->,very thick,gray] (0,4)--(0,3)--(1,4)--(2,4)--(1,3)--(0,2)--(0,1)--(1,2)--(2,3)--(3,4)--(4,4)--(3,3)--(2,2)--(1,1)--(0,0)--(1,0)--(2,1)--(3,2)--(4,3)--(4,2)--(3,1)--(2,0)--(3,0)--(4,1)--(4,0);
% %\node[fill=white] at (0,4.3){START};
% %\node[fill=white] at (4,-.3){STOP};
% \end{scope}

\begin{scope}[xshift=0cm,yshift=0cm]
\fill[thick,fill opacity=0.18,fill=gray] (0,0)--(1,0)--(1,1)--(0,1)--(0,0);
\end{scope}

\begin{scope}[xshift=0cm,yshift=2cm]
\fill[thick,fill opacity=0.18,fill=gray] (0,0)--(1,0)--(1,1)--(0,1)--(0,0);
\end{scope}

\begin{scope}[xshift=0cm,yshift=4cm]
\fill[thick,fill opacity=0.18,fill=gray] (0,0)--(1,0)--(1,1)--(0,1)--(0,0);
\end{scope}

\begin{scope}[xshift=1cm,yshift=-1cm]
\fill[thick,fill opacity=0.18,fill=gray] (0,0)--(1,0)--(1,1)--(0,1)--(0,0);
\end{scope}

\begin{scope}[xshift=1cm,yshift=1cm]
\fill[thick,fill opacity=0.18,fill=gray] (0,0)--(1,0)--(1,1)--(0,1)--(0,0);
\end{scope}

\begin{scope}[xshift=1cm,yshift=3cm]
\fill[thick,fill opacity=0.18,fill=gray] (0,0)--(1,0)--(1,1)--(0,1)--(0,0);
\end{scope}

\begin{scope}[xshift=2cm,yshift=0cm]
\fill[thick,fill opacity=0.18,fill=gray] (0,0)--(1,0)--(1,1)--(0,1)--(0,0);
\end{scope}

\begin{scope}[xshift=2cm,yshift=2cm]
\fill[thick,fill opacity=0.18,fill=gray] (0,0)--(1,0)--(1,1)--(0,1)--(0,0);
\end{scope}

\begin{scope}[xshift=2cm,yshift=4cm]
\fill[thick,fill opacity=0.18,fill=gray] (0,0)--(1,0)--(1,1)--(0,1)--(0,0);
\end{scope}

\begin{scope}[xshift=3cm,yshift=-1cm]
\fill[thick,fill opacity=0.18,fill=gray] (0,0)--(1,0)--(1,1)--(0,1)--(0,0);
\end{scope}

\begin{scope}[xshift=3cm,yshift=1cm]
\fill[thick,fill opacity=0.18,fill=gray] (0,0)--(1,0)--(1,1)--(0,1)--(0,0);
\end{scope}

\begin{scope}[xshift=3cm,yshift=3cm]
\fill[thick,fill opacity=0.18,fill=gray] (0,0)--(1,0)--(1,1)--(0,1)--(0,0);
\end{scope}

% \begin{scope}[xshift=6cm]
% \draw [->,very thick,gray] (0,4)--(0,3)--(1,4)--(2,4)--(1,3)--(0,2)--(0,1)--(1,2)--(2,3)--(3,4)--(4,4)--(3,3)--(2,2)--(1,1)--(0,0)--(1,0)--(2,1)--(3,2)--(4,3)--(4,2)--(3,1)--(2,0)--(3,0)--(4,1)--(4,0);
%\node[fill=white] at (0,4.3){START};
%\node[fill=white] at (4,-.3){STOP};
% \end{scope}

%------------------------------------------------------------------

\begin{scope}[xshift=0cm,yshift=3cm]
\node[text=black] at (0.5,0.5){2};
\end{scope}

\begin{scope}[xshift=-1cm,yshift=2cm]
\node[text=black] at (0.5,0.5){3};
\end{scope}

\begin{scope}[xshift=2cm,yshift=3cm]
\node[text=black] at (0.5,0.5){7};
\end{scope}

\begin{scope}[xshift=1cm,yshift=2cm]
\node[text=black] at (0.5,0.5){8};
\end{scope}

\begin{scope}[xshift=0cm,yshift=1cm]
\node[text=black] at (0.5,0.5){9};
\end{scope}

\begin{scope}[xshift=-1cm,yshift=0cm]
\node[text=black] at (0.5,0.5){10};
\end{scope}

\begin{scope}[xshift=4cm,yshift=3cm]
\node[text=black] at (0.5,0.5){15};
\end{scope}

\begin{scope}[xshift=3cm,yshift=2cm]
\node[text=black] at (0.5,0.5){16};
\end{scope}

\begin{scope}[xshift=2cm,yshift=1cm]
\node[text=black] at (0.5,0.5){17};
\end{scope}

\begin{scope}[xshift=1cm,yshift=0cm]
\node[text=black] at (0.5,0.5){18};
\end{scope}

\begin{scope}[xshift=4cm,yshift=1cm]
\node[text=black] at (0.5,0.5){22};
\end{scope}

\begin{scope}[xshift=3cm,yshift=0cm]
\node[text=black] at (0.5,0.5){23};
\end{scope}
%---------------------------------------

\begin{scope}[xshift=0cm,yshift=4cm]
\node[text=black] at (0.5,0.5){1};
\end{scope}

\begin{scope}[xshift=0cm,yshift=2cm]
\node[text=black] at (0.5,0.5){4};
\end{scope}

\begin{scope}[xshift=1cm,yshift=3cm]
\node[text=black] at (0.5,0.5){5};
\end{scope}

\begin{scope}[xshift=2cm,yshift=4cm]
\node[text=black] at (0.5,0.5){6};
\end{scope}

\begin{scope}[xshift=0cm,yshift=0cm]
\node[text=black] at (0.5,0.5){11};
\end{scope}

\begin{scope}[xshift=1cm,yshift=1cm]
\node[text=black] at (0.5,0.5){12};
\end{scope}

\begin{scope}[xshift=2cm,yshift=2cm]
\node[text=black] at (0.5,0.5){13};
\end{scope}

\begin{scope}[xshift=3cm,yshift=3cm]
\node[text=black] at (0.5,0.5){14};
\end{scope}

\begin{scope}[xshift=1cm,yshift=-1cm]
\node[text=black] at (0.5,0.5){19};
\end{scope}

\begin{scope}[xshift=2cm,yshift=0cm]
\node[text=black] at (0.5,0.5){20};
\end{scope}

\begin{scope}[xshift=3cm,yshift=1cm]
\node[text=black] at (0.5,0.5){21};
\end{scope}

\begin{scope}[xshift=3cm,yshift=-1cm]
\node[text=black] at (0.5,0.5){24};
\end{scope}

\begin{scope}
\node[circle, fill=black!75, inner sep=2pt, text=white, minimum size=10pt] at (0,4){1};
\node[circle, fill=black!75, inner sep=2pt, text=white, minimum size=10pt] at (1,4){3};
\node[circle, fill=black!75, inner sep=2pt, text=white, minimum size=10pt] at (2,4){4};
\node[circle, fill=black!75, inner sep=.5pt, text=white, minimum size=10pt] at (3,4){10};
\node[circle, fill=black!75, inner sep=.5pt, text=white, minimum size=10pt] at (4,4){11};
\node[circle, fill=black!75, inner sep=2pt, text=white, minimum size=10pt] at (0,3){2};
\node[circle, fill=black!75, inner sep=2pt, text=white, minimum size=10pt] at (1,3){5};
\node[circle, fill=black!75, inner sep=2pt, text=white, minimum size=10pt] at (2,3){9};
\node[circle, fill=black!75, inner sep=.5pt, text=white, minimum size=10pt] at (3,3){12};
\node[circle, fill=black!75, inner sep=.5pt, text=white, minimum size=10pt] at (4,3){19};
\node[circle, fill=black!75, inner sep=2pt, text=white, minimum size=10pt] at (0,2){6};
\node[circle, fill=black!75, inner sep=2pt, text=white, minimum size=10pt] at (1,2){8};
\node[circle, fill=black!75, inner sep=.5pt, text=white, minimum size=10pt] at (2,2){13};
\node[circle, fill=black!75, inner sep=.5pt, text=white, minimum size=10pt] at (3,2){18};
\node[circle, fill=black!75, inner sep=.5pt, text=white, minimum size=10pt] at (4,2){20};
\node[circle, fill=black!75, inner sep=2pt, text=white, minimum size=10pt] at (0,1){7};
\node[circle, fill=black!75, inner sep=.5pt, text=white, minimum size=10pt] at (1,1){14};
\node[circle, fill=black!75, inner sep=.5pt, text=white, minimum size=10pt] at (2,1){17};
\node[circle, fill=black!75, inner sep=.5pt, text=white, minimum size=10pt] at (3,1){21};
\node[circle, fill=black!75, inner sep=.5pt, text=white, minimum size=10pt] at (4,1){24};
\node[circle, fill=black!75, inner sep=.5pt, text=white, minimum size=10pt] at (0,0){15};
\node[circle, fill=black!75, inner sep=.5pt, text=white, minimum size=10pt] at (1,0){16};
\node[circle, fill=black!75, inner sep=.5pt, text=white, minimum size=10pt] at (2,0){22};
\node[circle, fill=black!75, inner sep=.5pt, text=white, minimum size=10pt] at (3,0){23};
\node[circle, fill=black!75, inner sep=.5pt, text=white, minimum size=10pt] at (4,0){25};
\end{scope}
% \begin{scope}[xshift=6cm]
% \draw [->,very thick] (0,4)--(0,3)--(1,4)--(2,4)--(1,3)--(0,2)--(0,1)--(1,2)--(2,3)--(3,4)--(4,4)--(3,3)--(2,2)--(1,1)--(0,0)--(1,0)--(2,1)--(3,2)--(4,3)--(4,2)--(3,1)--(2,0)--(3,0)--(4,1)--(4,0);
\end{tikzpicture}

%% file: figures/XZZXschedulegraph.tex
\begin{tikzpicture}[yscale=.5,xscale=.5]

\draw[xstep=1,ystep=1,lightgray,thin] (1,1) grid (24,25);

\node[rotate=45](A) at (20,16){$t=0,x=4$};
\draw (10.2,6.2)--(A);%(19.7,15.7);

\node[rotate=45](B) at (16.5,13.5) {$t=1,x=3$};
\draw (6.2,3.2)--(B)--(21.8,18.8);

\node[rotate=45](C) at (16,14){$t=2,x=2$};
\draw (3.2,1.2)--(C)--(23.8,21.8);

\node[rotate=45](D) at  (15.5,14.5) {$t=3,x=1$};
\draw (2.2,1.2)--(D)--(24,23);

\node[rotate=45](E) at (15,15) {$t=4,x=0$};
\draw (1,1)--(E)--(23.8,23.8);

\node[rotate=45](F) at (14.5,15.5) {$t=5,x=-1$};
\draw (1.2,2.2)--(F)--(24,25);

\node[rotate=45](G) at (14,16) {$t=6,x=-2$};
\draw (1,3)--(G)--(23,25);

\node[rotate=45](H) at (13.5,16.5) {$t=7,x=-3$};
\draw (1.2,4.2)--(H)--(21.8,24.8);

\node[rotate=45](I) at (13,17) {$t=8,x=-4$};
\draw (2.2,6.2)--(I)--(18.8,22.8);

\node[rotate=45](J) at (16,21) {$t=9,x=-5$};
\draw (6.2,11.2)--(J);

\node[circle, fill=black!75, inner sep=.5pt, text=white, minimum size=10pt,name=r,label={[shift=(r.west)]left:1}] at (0,1){};
\node[circle, fill=black!75, inner sep=.5pt, text=white, minimum size=10pt,name=r,label={[shift=(r.west)]left:2}] at (0,2){};
\node[circle, fill=black!75, inner sep=.5pt, text=white, minimum size=10pt,name=r,label={[shift=(r.west)]left:3}] at (0,3){};
\node[circle, fill=black!75, inner sep=.5pt, text=white, minimum size=10pt,name=r,label={[shift=(r.west)]left:4}] at (0,4){};
\node[circle, fill=black!75, inner sep=.5pt, text=white, minimum size=10pt,name=r,label={[shift=(r.west)]left:5}] at (0,5){};
\node[circle, fill=black!75, inner sep=.5pt, text=white, minimum size=10pt,name=r,label={[shift=(r.west)]left:6}] at (0,6){};
\node[circle, fill=black!75, inner sep=.5pt, text=white, minimum size=10pt,name=r,label={[shift=(r.west)]left:7}] at (0,7){};
\node[circle, fill=black!75, inner sep=.5pt, text=white, minimum size=10pt,name=r,label={[shift=(r.west)]left:8}] at (0,8){};
\node[circle, fill=black!75, inner sep=.5pt, text=white, minimum size=10pt,name=r,label={[shift=(r.west)]left:9}] at (0,9){};
\node[circle, fill=black!75, inner sep=.5pt, text=white, minimum size=10pt,name=r,label={[shift=(r.west)]left:10}] at (0,10){};
\node[circle, fill=black!75, inner sep=.5pt, text=white, minimum size=10pt,name=r,label={[shift=(r.west)]left:11}] at (0,11){};
\node[circle, fill=black!75, inner sep=.5pt, text=white, minimum size=10pt,name=r,label={[shift=(r.west)]left:12}] at (0,12){};
\node[circle, fill=black!75, inner sep=.5pt, text=white, minimum size=10pt,name=r,label={[shift=(r.west)]left:13}] at (0,13){};
\node[circle, fill=black!75, inner sep=.5pt, text=white, minimum size=10pt,name=r,label={[shift=(r.west)]left:14}] at (0,14){};
\node[circle, fill=black!75, inner sep=.5pt, text=white, minimum size=10pt,name=r,label={[shift=(r.west)]left:15}] at (0,15){};
\node[circle, fill=black!75, inner sep=.5pt, text=white, minimum size=10pt,name=r,label={[shift=(r.west)]left:16}] at (0,16){};
\node[circle, fill=black!75, inner sep=.5pt, text=white, minimum size=10pt,name=r,label={[shift=(r.west)]left:17}] at (0,17){};
\node[circle, fill=black!75, inner sep=.5pt, text=white, minimum size=10pt,name=r,label={[shift=(r.west)]left:18}] at (0,18){};
\node[circle, fill=black!75, inner sep=.5pt, text=white, minimum size=10pt,name=r,label={[shift=(r.west)]left:19}] at (0,19){};
\node[circle, fill=black!75, inner sep=.5pt, text=white, minimum size=10pt,name=r,label={[shift=(r.west)]left:20}] at (0,20){};
\node[circle, fill=black!75, inner sep=.5pt, text=white, minimum size=10pt,name=r,label={[shift=(r.west)]left:21}] at (0,21){};
\node[circle, fill=black!75, inner sep=.5pt, text=white, minimum size=10pt,name=r,label={[shift=(r.west)]left:22}] at (0,22){};
\node[circle, fill=black!75, inner sep=.5pt, text=white, minimum size=10pt,name=r,label={[shift=(r.west)]left:23}] at (0,23){};
\node[circle, fill=black!75, inner sep=.5pt, text=white, minimum size=10pt,name=r,label={[shift=(r.west)]left:24}] at (0,24){};
\node[circle, fill=black!75, inner sep=.5pt, text=white, minimum size=10pt,name=r,label={[shift=(r.west)]left:25}] at (0,25){};

\node[circle, draw,fill opacity=0.18,fill=gray, inner sep=.5pt, text=white, minimum size=10pt,name=r,label={[shift=(r.south)]below:1}] at (1,0){};
\node[circle, draw,fill opacity=0.3,fill=darkgray, inner sep=.5pt, text=white, minimum size=10pt,name=r,label={[shift=(r.south)]below:2}] at (2,0){};
\node[circle, draw,fill opacity=0.3,fill=darkgray, inner sep=.5pt, text=white, minimum size=10pt,name=r,label={[shift=(r.south)]below:3}] at (3,0){};
\node[circle, draw,fill opacity=0.18,fill=gray, inner sep=.5pt, text=white, minimum size=10pt,name=r,label={[shift=(r.south)]below:4}] at (4,0){};
\node[circle, draw,fill opacity=0.18,fill=gray, inner sep=.5pt, text=white, minimum size=10pt,name=r,label={[shift=(r.south)]below:5}] at (5,0){};
\node[circle, draw,fill opacity=0.18,fill=gray, inner sep=.5pt, text=white, minimum size=10pt,name=r,label={[shift=(r.south)]below:6}] at (6,0){};
\node[circle, draw,fill opacity=0.3,fill=darkgray, inner sep=.5pt, text=white, minimum size=10pt,name=r,label={[shift=(r.south)]below:7}] at (7,0){};
\node[circle, draw,fill opacity=0.3,fill=darkgray, inner sep=.5pt, text=white, minimum size=10pt,name=r,label={[shift=(r.south)]below:8}] at (8,0){};
\node[circle, draw,fill opacity=0.3,fill=darkgray, inner sep=.5pt, text=white, minimum size=10pt,name=r,label={[shift=(r.south)]below:9}] at (9,0){};
\node[circle, draw,fill opacity=0.3,fill=darkgray, inner sep=.5pt, text=white, minimum size=10pt,name=r,label={[shift=(r.south)]below:10}] at (10,0){};
\node[circle, draw,fill opacity=0.18,fill=gray, inner sep=.5pt, text=white, minimum size=10pt,name=r,label={[shift=(r.south)]below:11}] at (11,0){};
\node[circle, draw,fill opacity=0.18,fill=gray, inner sep=.5pt, text=white, minimum size=10pt,name=r,label={[shift=(r.south)]below:12}] at (12,0){};
\node[circle, draw,fill opacity=0.18,fill=gray, inner sep=.5pt, text=white, minimum size=10pt,name=r,label={[shift=(r.south)]below:13}] at (13,0){};
\node[circle, draw,fill opacity=0.18,fill=gray, inner sep=.5pt, text=white, minimum size=10pt,name=r,label={[shift=(r.south)]below:14}] at (14,0){};
\node[circle, draw,fill opacity=0.3,fill=darkgray, inner sep=.5pt, text=white, minimum size=10pt,name=r,label={[shift=(r.south)]below:15}] at (15,0){};
\node[circle, draw,fill opacity=0.3,fill=darkgray, inner sep=.5pt, text=white, minimum size=10pt,name=r,label={[shift=(r.south)]below:16}] at (16,0){};
\node[circle, draw,fill opacity=0.3,fill=darkgray, inner sep=.5pt, text=white, minimum size=10pt,name=r,label={[shift=(r.south)]below:17}] at (17,0){};
\node[circle, draw,fill opacity=0.3,fill=darkgray, inner sep=.5pt, text=white, minimum size=10pt,name=r,label={[shift=(r.south)]below:18}] at (18,0){};
\node[circle, draw,fill opacity=0.18,fill=gray, inner sep=.5pt, text=white, minimum size=10pt,name=r,label={[shift=(r.south)]below:19}] at (19,0){};
\node[circle, draw,fill opacity=0.18,fill=gray, inner sep=.5pt, text=white, minimum size=10pt,name=r,label={[shift=(r.south)]below:20}] at (20,0){};
\node[circle, draw,fill opacity=0.18,fill=gray, inner sep=.5pt, text=white, minimum size=10pt,name=r,label={[shift=(r.south)]below:21}] at (21,0){};
\node[circle, draw,fill opacity=0.3,fill=darkgray, inner sep=.5pt, text=white, minimum size=10pt,name=r,label={[shift=(r.south)]below:22}] at (22,0){};
\node[circle, draw,fill opacity=0.3,fill=darkgray, inner sep=.5pt, text=white, minimum size=10pt,name=r,label={[shift=(r.south)]below:23}] at (23,0){};
\node[circle, draw,fill opacity=0.18,fill=gray, inner sep=.5pt, text=white, minimum size=10pt,name=r,label={[shift=(r.south)]below:24}] at (24,0){};

\node[circle, fill=darkergreen, inner sep=.5pt, text=white, minimum size=7pt] at (1,1){};
\node[circle, fill=darkergreen, inner sep=.5pt, text=white, minimum size=7pt] at (1,3){};
\node[circle, fill=darkergreen, inner sep=.5pt, text=white, minimum size=7pt] at (2,1){};
\node[circle, fill=darkergreen, inner sep=.5pt, text=white, minimum size=7pt] at (2,2){};
\node[circle, fill=darkergreen, inner sep=.5pt, text=white, minimum size=7pt] at (2,3){};
\node[circle, fill=darkergreen, inner sep=.5pt, text=white, minimum size=7pt] at (2,5){};
\node[circle, fill=darkergreen, inner sep=.5pt, text=white, minimum size=7pt] at (3,2){};
\node[circle, fill=darkergreen, inner sep=.5pt, text=white, minimum size=7pt] at (3,6){};
\node[circle, fill=darkergreen, inner sep=.5pt, text=white, minimum size=7pt] at (4,2){};
\node[circle, fill=darkergreen, inner sep=.5pt, text=white, minimum size=7pt] at (4,5){};
\node[circle, fill=darkergreen, inner sep=.5pt, text=white, minimum size=7pt] at (4,6){};
\node[circle, fill=darkergreen, inner sep=.5pt, text=white, minimum size=7pt] at (4,8){};
\node[circle, fill=darkergreen, inner sep=.5pt, text=white, minimum size=7pt] at (5,3){};
\node[circle, fill=darkergreen, inner sep=.5pt, text=white, minimum size=7pt] at (5,4){};
\node[circle, fill=darkergreen, inner sep=.5pt, text=white, minimum size=7pt] at (5,5){};
\node[circle, fill=darkergreen, inner sep=.5pt, text=white, minimum size=7pt] at (5,9){};
\node[circle, fill=darkergreen, inner sep=.5pt, text=white, minimum size=7pt] at (6,4){};
\node[circle, fill=darkergreen, inner sep=.5pt, text=white, minimum size=7pt] at (6,10){};
\node[circle, fill=darkergreen, inner sep=.5pt, text=white, minimum size=7pt] at (7,4){};
\node[circle, fill=darkergreen, inner sep=.5pt, text=white, minimum size=7pt] at (7,9){};
\node[circle, fill=darkergreen, inner sep=.5pt, text=white, minimum size=7pt] at (7,10){};
\node[circle, fill=darkergreen, inner sep=.5pt, text=white, minimum size=7pt] at (7,12){};
\node[circle, fill=darkergreen, inner sep=.5pt, text=white, minimum size=7pt] at (8,5){};
\node[circle, fill=darkergreen, inner sep=.5pt, text=white, minimum size=7pt] at (8,8){};
\node[circle, fill=darkergreen, inner sep=.5pt, text=white, minimum size=7pt] at (8,9){};
\node[circle, fill=darkergreen, inner sep=.5pt, text=white, minimum size=7pt] at (8,13){};
\node[circle, fill=darkergreen, inner sep=.5pt, text=white, minimum size=7pt] at (9,6){};
\node[circle, fill=darkergreen, inner sep=.5pt, text=white, minimum size=7pt] at (9,7){};
\node[circle, fill=darkergreen, inner sep=.5pt, text=white, minimum size=7pt] at (9,8){};
\node[circle, fill=darkergreen, inner sep=.5pt, text=white, minimum size=7pt] at (9,14){};
\node[circle, fill=darkergreen, inner sep=.5pt, text=white, minimum size=7pt] at (10,7){};
\node[circle, fill=darkergreen, inner sep=.5pt, text=white, minimum size=7pt] at (10,15){};
\node[circle, fill=darkergreen, inner sep=.5pt, text=white, minimum size=7pt] at (11,7){};
\node[circle, fill=darkergreen, inner sep=.5pt, text=white, minimum size=7pt] at (11,14){};
\node[circle, fill=darkergreen, inner sep=.5pt, text=white, minimum size=7pt] at (11,15){};
\node[circle, fill=darkergreen, inner sep=.5pt, text=white, minimum size=7pt] at (11,16){};
\node[circle, fill=darkergreen, inner sep=.5pt, text=white, minimum size=7pt] at (12,8){};
\node[circle, fill=darkergreen, inner sep=.5pt, text=white, minimum size=7pt] at (12,13){};
\node[circle, fill=darkergreen, inner sep=.5pt, text=white, minimum size=7pt] at (12,14){};
\node[circle, fill=darkergreen, inner sep=.5pt, text=white, minimum size=7pt] at (12,17){};
\node[circle, fill=darkergreen, inner sep=.5pt, text=white, minimum size=7pt] at (13,9){};
\node[circle, fill=darkergreen, inner sep=.5pt, text=white, minimum size=7pt] at (13,12){};
\node[circle, fill=darkergreen, inner sep=.5pt, text=white, minimum size=7pt] at (13,13){};
\node[circle, fill=darkergreen, inner sep=.5pt, text=white, minimum size=7pt] at (13,18){};
\node[circle, fill=darkergreen, inner sep=.5pt, text=white, minimum size=7pt] at (14,10){};
\node[circle, fill=darkergreen, inner sep=.5pt, text=white, minimum size=7pt] at (14,11){};
\node[circle, fill=darkergreen, inner sep=.5pt, text=white, minimum size=7pt] at (14,12){};
\node[circle, fill=darkergreen, inner sep=.5pt, text=white, minimum size=7pt] at (14,19){};
\node[circle, fill=darkergreen, inner sep=.5pt, text=white, minimum size=7pt] at (15,11){};
\node[circle, fill=darkergreen, inner sep=.5pt, text=white, minimum size=7pt] at (15,19){};
\node[circle, fill=darkergreen, inner sep=.5pt, text=white, minimum size=7pt] at (16,12){};
\node[circle, fill=darkergreen, inner sep=.5pt, text=white, minimum size=7pt] at (16,18){};
\node[circle, fill=darkergreen, inner sep=.5pt, text=white, minimum size=7pt] at (16,19){};
\node[circle, fill=darkergreen, inner sep=.5pt, text=white, minimum size=7pt] at (16,20){};
\node[circle, fill=darkergreen, inner sep=.5pt, text=white, minimum size=7pt] at (17,13){};
\node[circle, fill=darkergreen, inner sep=.5pt, text=white, minimum size=7pt] at (17,17){};
\node[circle, fill=darkergreen, inner sep=.5pt, text=white, minimum size=7pt] at (17,18){};
\node[circle, fill=darkergreen, inner sep=.5pt, text=white, minimum size=7pt] at (17,21){};
\node[circle, fill=darkergreen, inner sep=.5pt, text=white, minimum size=7pt] at (18,14){};
\node[circle, fill=darkergreen, inner sep=.5pt, text=white, minimum size=7pt] at (18,16){};
\node[circle, fill=darkergreen, inner sep=.5pt, text=white, minimum size=7pt] at (18,17){};
\node[circle, fill=darkergreen, inner sep=.5pt, text=white, minimum size=7pt] at (18,22){};
\node[circle, fill=darkergreen, inner sep=.5pt, text=white, minimum size=7pt] at (19,16){};
\node[circle, fill=darkergreen, inner sep=.5pt, text=white, minimum size=7pt] at (19,22){};
\node[circle, fill=darkergreen, inner sep=.5pt, text=white, minimum size=7pt] at (20,17){};
\node[circle, fill=darkergreen, inner sep=.5pt, text=white, minimum size=7pt] at (20,21){};
\node[circle, fill=darkergreen, inner sep=.5pt, text=white, minimum size=7pt] at (20,22){};
\node[circle, fill=darkergreen, inner sep=.5pt, text=white, minimum size=7pt] at (20,23){};
\node[circle, fill=darkergreen, inner sep=.5pt, text=white, minimum size=7pt] at (21,18){};
\node[circle, fill=darkergreen, inner sep=.5pt, text=white, minimum size=7pt] at (21,20){};
\node[circle, fill=darkergreen, inner sep=.5pt, text=white, minimum size=7pt] at (21,21){};
\node[circle, fill=darkergreen, inner sep=.5pt, text=white, minimum size=7pt] at (21,24){};
\node[circle, fill=darkergreen, inner sep=.5pt, text=white, minimum size=7pt] at (22,20){};
\node[circle, fill=darkergreen, inner sep=.5pt, text=white, minimum size=7pt] at (22,24){};
\node[circle, fill=darkergreen, inner sep=.5pt, text=white, minimum size=7pt] at (23,21){};
\node[circle, fill=darkergreen, inner sep=.5pt, text=white, minimum size=7pt] at (23,23){};
\node[circle, fill=darkergreen, inner sep=.5pt, text=white, minimum size=7pt] at (23,24){};
\node[circle, fill=darkergreen, inner sep=.5pt, text=white, minimum size=7pt] at (23,25){};
\node[circle, fill=darkergreen, inner sep=.5pt, text=white, minimum size=7pt] at (24,23){};
\end{tikzpicture}